\documentclass[12pt,a4paper]{article}

\usepackage[nosort]{cite}
\usepackage[hyperref,bulletsep]{collect}
\usepackage{graphicx}
\usepackage{bbm}
\usepackage{amsmath,amssymb}
\usepackage{subfigure}
\usepackage{verbatim}

\makeatletter \@addtoreset{equation}{section} \makeatother

\makeatletter
\let\old@startsection=\@startsection
\let\oldl@section=\l@section
\renewcommand{\@startsection}[6]{\old@startsection{#1}{#2}{#3}{#4}{#5}{#6\mathversion{bold}}}
\renewcommand{\l@section}[2]{\oldl@section{\mathversion{bold}#1}{#2}}
\makeatother

\makeatletter
\let\old@makecaption=\@makecaption
\def\@makecaption{\small\old@makecaption}
\makeatother

\renewcommand{\thefootnote}{\arabic{footnote}}
\let\oldPhi=\Phi
\let\oldPsi=\Psi
\let\oldGamma=\Gamma
\let\oldDelta=\Delta
\let\oldSigma=\Sigma
\let\oldTheta=\Theta
\let\oldPi=\Pi
\let\oldUpsilon=\Upsilon
\renewcommand{\Phi}{\mathnormal{\oldPhi}}
\renewcommand{\Psi}{\mathnormal{\oldPsi}}
\renewcommand{\Gamma}{\mathnormal{\oldGamma}}
\renewcommand{\Sigma}{\mathnormal{\oldSigma}}
\renewcommand{\Delta}{\mathnormal{\oldDelta}}
\renewcommand{\Theta}{\mathnormal{\oldTheta}}
\renewcommand{\Pi}{\mathnormal{\oldPi}}
\renewcommand{\Upsilon}{\mathnormal{\oldUpsilon}}

\newcommand{\superN}{\mathcal{N}}
\newcommand{\Action}{\mathcal{S}}
\newcommand{\Lagr}{\mathcal{L}}

\newcommand{\sign}{\mathop{\mathrm{sign}}}

\newcommand{\order}{\mathcal{O}}

\newcommand{\Reals}{\mathbbm{R}}

\newcommand{\Sphere}{S}  % {\mathbbm{S}}
\newcommand{\AdS}{\mathrm{AdS}}

\ifx\genfrac\sdflkaj

\else

\fi
\newcommand{\sfrac}[2]{{\textstyle\frac{#1}{#2}}}
\newcommand{\half}{\sfrac{1}{2}}
\newcommand{\ihalf}{\sfrac{i}{2}}

\newcommand{\rep}[1]{{\mathbf{#1}}}

\newcommand{\alg}[1]{\mathfrak{#1}}
\newcommand{\grp}[1]{\mathrm{#1}}

\newcommand{\grSU}{\grp{SU}}
\newcommand{\grSO}{\grp{SO}}

\newcommand{\algSU}{\alg{su}}
\newcommand{\algSO}{\alg{so}}
\newcommand{\algSL}{\alg{sl}}
\newcommand{\algPSU}{\alg{psu}}

\newcommand{\brk}[1]{(#1)}
\newcommand{\lrbrk}[1]{\left(#1\right)}
\newcommand{\bigbrk}[1]{\bigl(#1\bigr)}
\newcommand{\Bigbrk}[1]{\Bigl(#1\Bigr)}
\newcommand{\biggbrk}[1]{\biggl(#1\biggr)}

\newcommand{\lrsbrk}[1]{\left[#1\right]}
\newcommand{\bigsbrk}[1]{\bigl[#1\bigr]}
\newcommand{\Bigsbrk}[1]{\Bigl[#1\Bigr]}
\newcommand{\biggsbrk}[1]{\biggl[#1\biggr]}

\newcommand{\ket}[1]{\mathopen{|}#1\mathclose{\rangle}}
\newcommand{\bra}[1]{\mathopen{\langle}#1\mathclose{|}}
\newcommand{\braket}[2]{\mathopen{\langle}#1|#2\mathclose{\rangle}}

\newcommand{\comm}[2]{[#1,#2]}

\newcommand{\lrabs}[1]{\left|#1\right|}
\newcommand{\abs}[1]{{|#1|}}

\newcommand{\bigeval}[1]{#1\big|}

\def\[{\begin{equation}}
\def\]{\end{equation}}
\newcommand{\be}{\begin{eqnarray}}
\newcommand{\ee}{\end{eqnarray}}
\newcommand{\nn}{\nonumber}
\newcommand{\nln}{\nonumber\\}
\newcommand{\nl}[1][0pt]{\nonumber\\[#1]&\hspace{-4\arraycolsep}&\mathord{}}

\newcommand{\earel}[1]{\mathrel{}&\hspace{-2\arraycolsep}#1\hspace{-2\arraycolsep}&\mathrel{}}
\newcommand{\eq}{\earel{=}}

\makeatletter
\def\mr@ignsp#1 {\ifx\:#1\@empty\else #1\expandafter\mr@ignsp\fi}%
\newcommand{\multiref}[1]{\begingroup%\let\protect\string%
\xdef\mr@no@sparg{\expandafter\mr@ignsp#1 \: }%
\def\mr@comma{}%
\@for\mr@refs:=\mr@no@sparg\do{\mr@comma\def\mr@comma{,}\ref{\mr@refs}}%
\endgroup}
\makeatother

\newcommand{\hypref}[2]{\ifx\href\asklfhas #2\else\href{#1}{#2}\fi}

\newcommand{\secref}[1]{Sec.~\multiref{#1}}

\newcommand{\appref}[1]{App.~\multiref{#1}}

\newcommand{\tabref}[1]{Tab.~\multiref{#1}}

\newcommand{\figref}[1]{Fig.~\multiref{#1}}
\renewcommand{\eqref}[1]{(\multiref{#1})}

\ifx\href\asklfhas\newcommand{\href}[2]{#2}\fi

\newcommand{\Smatrix}{S}  % S-matrix for full theory
\newcommand{\smatrix}{\mathsf{S}}          % S-matrix for one su(2|2) factor
\newcommand{\Amp}{\mathcal{A}}      % amplitude
\newcommand{\Samp}{\mathcal{S}}
\newcommand{\Ramp}{\mathcal{R}}
\newcommand{\lAA}{{a}}
\newcommand{\rAA}{{\dot{a}}}
\newcommand{\laa}{{\alpha}}
\newcommand{\raa}{{\dot{\alpha}}}
\newcommand{\lBB}{{b}}

\newcommand{\lbb}{{\beta}}

\newcommand{\lCC}{{c}}

\newcommand{\lcc}{{\gamma}}

\newcommand{\lDD}{{d}}

\newcommand{\ldd}{{\delta}}

\newcommand{\comma}{\quad,\quad}
\newcommand{\unit}{\mathbbm{1}}
\newcommand{\tim}[1]{\dot{#1}}
\newcommand{\spa}[1]{\acute{#1}}
\newcommand{\Bsi}{\Upsilon}
\newcommand{\g}{\gamma}
\newcommand{\levi}{\epsilon}

\newcommand{\eps}{\varepsilon}
\renewcommand{\vec}[1]{\mathbf{#1}}
\newcommand{\pin}{\eta}                  % the sequence (a,b,c,d,e,f)
\newcommand{\pou}{\bar{\eta}}            % the sequence (a,b,c,-d,-e,-f)
\newcommand{\vecpin}{\boldsymbol{\eta}}

\newcommand{\vecsigma}{\boldsymbol{\sigma}}

\newcommand{\prop}{\mathsf{I}}
\newcommand{\dog}{\mathsf{D}}
\newcommand{\sun}{\mathsf{O}}
\newcommand{\bubble}{\mathsf{B}}

\begin{document}

\thispagestyle{empty}
\begin{flushright}\footnotesize
\texttt{arxiv:0707.2082}\\
\texttt{UUITP-14/07} \vspace{0.8cm}
\end{flushright}

\renewcommand{\thefootnote}{\fnsymbol{footnote}}
\setcounter{footnote}{0}

\begin{center}
{\Large\textbf{\mathversion{bold}
Factorized world-sheet scattering \\
in near-flat $\AdS_5\times\Sphere^5$
}\par}

\vspace{1.5cm}

\textrm{Valentina Giangreco M.~Puletti, Thomas Klose and Olof Ohlsson Sax} \vspace{8mm}

\textit{
Department of Theoretical Physics, Uppsala University\\
SE-75108 Uppsala, Sweden}\\
\texttt{Valentina.Giangreco,Thomas.Klose,Olof.Ohlsson-Sax@teorfys.uu.se} \vspace{3mm}

%%%%%%%%

\par\vspace{1cm}

\textbf{Abstract} \vspace{5mm}

\begin{minipage}{13cm}
We show the factorization of the three-particle world-sheet S-matrix of $\AdS_5\times\Sphere^5$ superstring theory in the near-flat-space limit at one loop order. This is done by computing various scattering amplitudes from Feynman diagrams in the world-sheet theory. The knowledge of certain highest weight amplitudes allows us to fix all the freedom in the three-particle S-matrix, which we argue to be constrained up to four scalar functions due to the symmetries of the model. We demonstrate that these amplitudes are given by corresponding products of the known two-particle S-matrix elements, from which it follows that the scattering of any three world-sheet excitations factorizes. This provides an explicit and direct check of the quantum integrability of string theory in near-flat $\AdS_5\times\Sphere^5$ as it renders evidence for the existence of higher conserved charges. By computing further amplitudes we also obtain an indirect confirmation of the supersymmetries of the near-flat-space model.
\end{minipage}

\end{center}

\vspace{0.5cm}

%%%%%%%%%%%%%%%%%%%%%%%%%%%%%%%%%%%%%%%%%%%%%%%%%%%%%%%%%%%%%%%%%%%%%%%%%%%
\newpage
\setcounter{page}{1}
\renewcommand{\thefootnote}{\arabic{footnote}}
\setcounter{footnote}{0}
\tableofcontents

%%%%%%%%%%%%%%%%%%%%%%%%%%%%%%%%%%%%%%%%%%%%%%%%%%%%%%%%%%%%%%%%%%%%%%%%%%%
%%%%%%%%%%%%%%%%%%%%%%%%%%%%%%%%%%%%%%%%%%%%%%%%%%%%%%%%%%%%%%%%%%%%%%%%%%%
\section{Introduction}

The integrable structures found in type IIB superstring theory on $AdS_5 \times S^5$ \cite{Mandal:2002fs, Bena:2003wd, Kazakov:2004qf} and in $\superN=4$ Super Yang-Mills theory \cite{Minahan:2002ve, Beisert:2003tq, Beisert:2003yb} have turned the planar AdS/CFT correspondence \cite{Maldacena:1998re, Gubser:1998bc, Witten:1998qj} into a two-dimensional particle model%
\footnote{We refer to the review articles \cite{Beisert:2004ry, Zarembo:2004hp, Plefka:2005bk, Minahan:2006sk} for an introduction to integrability in AdS/CFT and also as guides to the vast amount of literature on this subject.}%
. At the two ends of the coupling range, these particles represent the excitations of the string world-sheet and the impurities in gauge invariant operators, respectively. The asymptotic AdS/CFT spectrum of long strings and heavy operators is then encoded in Bethe equations \cite{Beisert:2005fw} which derive from the S-matrix that describes the interactions of these particles \cite{Staudacher:2004tk, Beisert:2005tm, Arutyunov:2006yd}. Global symmetries fix the two-particle S-matrix up to an overall phase factor \cite{Beisert:2005tm}. This dressing phase embodies the remaining freedom which is needed for a smooth interpolation between the two sides of the correspondence. The phase is constrained by a crossing equation \cite{Janik:2006dc}, and based on \cite{Arutyunov:2004vx, Hernandez:2006tk, Beisert:2006ib} a non-perturbative expression was proposed in \cite{Beisert:2006ez}.

The role of integrability is to ensure that the multi-particle S-matrices factorize into products of two-particle S-matrix elements \cite{Kulish:1975ba, Luscher:1977uq, Shankar:1977cm, Zamolodchikov:1978xm}\footnote{For a very nice review on factorized scattering see \cite{Dorey:1996gd}.} such that knowing the latter is enough to determine the spectrum for states with an arbitrary number of excitations. In this context two things are currently the subject of extensive investigations: proving both the correctness of the phase factor and the persistence of integrability for all values of the coupling constant.

By now, the proposed two-particle S-matrix has been verified by computations of the cusp anomalous dimension at weak coupling up to fourth order \cite{Bern:2006ew,Cachazo:2006az} and at strong coupling up to first order \cite{Benna:2006nd, Alday:2007qf, Kostov:2007kx, Beccaria:2007tk, Casteill:2007ct}, while the two-loop calculation suffers from divergences \cite{Roiban:2007jf}. The phase factor was also directly verified by a gauge theory computation at four loops \cite{Beisert:2007hz} and string theory computations for light-cone modes up to two loops \cite{Klose:2007wq, Klose:2007rz} and giant magnons up to one loop \cite{Chen:2007vs}. Moreover, there are numerous check of the one-loop phase factor by comparing energy corrections to classical string configurations \cite{Frolov:2002av, Frolov:2003tu, Frolov:2004bh, Park:2005ji, Frolov:2006qe} to Bethe-ansatz predictions \cite{Schafer-Nameki:2005tn, Beisert:2005cw, Schafer-Nameki:2005is, Schafer-Nameki:2006gk, Freyhult:2006vr, Schafer-Nameki:2006ey}. The physical origin of the phase was discussed in \cite{Rej:2007vm}\cite{Sakai:2007rk} and a method to derive it from first principles was advertised in \cite{Gromov:2007cd}. Its analytical structure agrees with the expectations based on the exact spectrum \cite{Dorey:2007xn}.

The usefulness of the two-particle S-matrix, however, relies on the existence of exact integrability.\footnote{For completeness, we should mention that there are examples where the possession of the two-particle S-matrix does not guarantee the diagonalizability of the Hamiltonian of the system \cite{Das:2007tb}.} On the gauge theory side, integrability was proven at the leading order by means of an algebraic Bethe ansatz \cite{Minahan:2002ve, Beisert:2003yb}. On the string theory side, the classical integrability follows from the existence of infinitely many non-local charges \cite{Mandal:2002fs, Bena:2003wd}. For checking integrability beyond leading order there are various strategies: In gauge theory, the early attention has been directed towards the degeneracies in the quantum spectrum and the construction of responsible conserved charges \cite{Beisert:2003tq, Beisert:2003ys, Serban:2004jf, Beisert:2004hm}. A more algebraic approach is to understand higher order integrability in terms of Yangian symmetries \cite{Serban:2004jf, Agarwal:2004sz, Berkovits:2004xu, Zwiebel:2006cb, Beisert:2007ds}. A very direct check of quantum integrability is to show S-matrix factorization, performed on the gauge theory side for the three-particle S-matrix in the $\algSL(2)$ sector up to two-loops in \cite{Eden:2006rx}. In contrast to the fairly strong results for the gauge theory, the quantum integrability of the AdS string theory is still essentially conjectural. One possibility to judge about integrability beyond the classical level is to verify if the proposed quantum Bethe equations correctly predict the energies for states with at least three excitations. This has been successfully done in the near-plane-wave limit \cite{Callan:2004ev} and in the uniform light-cone gauge \cite{Klose:2006zd, Hentschel:2007xn}, but only semi-classically. Very recently, a stronger hint for quantum integrability was given by showing that all logarithmic divergences in the transfer matrix cancel \cite{Mikhailov:2007mr}.

This article provides the first direct check of \emph{quantum integrability} on the string theory side by focusing on the factorization of the S-matrix and the absence of particle production in the world-sheet theory. Since the full $AdS_5\times S^5$ sigma-model \cite{Metsaev:1998it} is quite complicated, we work in the near-flat-space limit \cite{Maldacena:2006rv, Benvenuti:2007qt}. This simplifies the sigma model immensely, but nonetheless permits probing the world-sheet scattering of the full theory as shown in the two-particle case \cite{Klose:2007wq, Klose:2007rz}. Here, we compute three-particle scattering amplitudes at one-loop order ($1/\sqrt{\lambda}\,{}^{3}$) and show that the S-matrix indeed factorizes. This implies the existence of at least one higher conserved charge and therefore supplies strong evidence for the quantum integrability \cite{Parke:1980ki}.

%%%%%%%%%%%%%%%%%%%%%%%%%%%%%%%%%%%%%%%%%%%%%%%%%%%%%%%%%%%%%%%%%%%%%%%%%%%
\paragraph{Outline.} In the remainder of this introduction, we recall some facts about factorized scattering, write down the precise factorization equation which we are going to prove, and sketch the mechanism of factorization from a Feynman diagram point of view. In \secref{sec:MS-model}, we briefly review the near-flat-space model and quote the world-sheet two-particle S-matrix in this limit. \secref{sec:symmetries} is devoted to the discussion of the symmetries of the S-matrix which reveals that the three-particle S-matrix has only four independent components. We compute these components in \secref{sec:3p-scatt} from Feynman diagrams to one-loop order. We also compute some further S-matrix elements as an indirect check of the symmetries. In \secref{sec:factorization}, we show that the three-particle S-matrix elements indeed factorize into products of elements of the two-particle S-matrix. We do this by computing the expected three-particle S-matrix from the two-particle S-matrix determined by the Zamolodchikov algebra and show that it agrees with the result of the Feynman diagram computation. On the basis of the symmetries of the model, this proves the factorization of the three-particle scattering in the $\AdS_5\times\Sphere^5$ superstring world-sheet theory in the near-flat limit to the first quantum level. In the appendices, we set up the notation (\appref{app:notation}), discuss the $\algSU(2)^4$ transformations of the fields (\appref{app:embedding}), describe the three-particle phase space (\appref{app:phasespace}) and give more details for the computation of the Feynman diagrams (\appref{app:diagrams}).

\begin{figure}
\begin{center}
\includegraphics[scale=0.6]{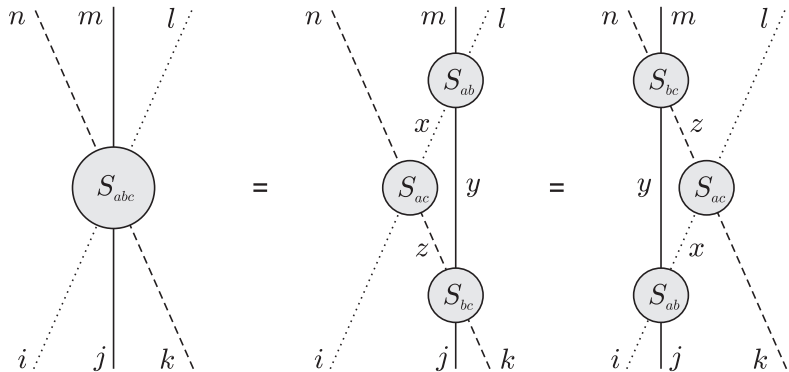}
\end{center}
\caption{\textbf{S-matrix factorization.} This graphic illustrates equation \protect\eqref{eqn:Smat-factorization}, the factorization of the three particle S-matrix into the product of three two-particle S-matrices. The dotted, solid and dashed lines represent the momenta $a$, $b$ and $c$, respectively. The other labels are the flavor indices of the fields. The S-matrices dictate the flavor indices how to hop between the momentum lines.}
\label{fig:factorization}
\end{figure}

%%%%%%%%%%%%%%%%%%%%%%%%%%%%%%%%%%%%%%%%%%%%%%%%%%%%%%%%%%%%%%%%%%%%%%%%%%%
\paragraph{Facts about factorized scattering.} In 1+1-dimensional field theories the existence of infinitely many conserved charges is equivalent to the factorization of the S-matrix \cite{Kulish:1975ba, Luscher:1977uq, Shankar:1977cm, Zamolodchikov:1978xm}. Hence, one can prove integrability by showing that all multi-particle S-matrices can be expressed in terms of the two-particle S-matrix according to the Zamolodchikov algebra \cite{Zamolodchikov:1978xm}. We record the precise relationship for the case of three-particle scattering, which is the case we will be dealing with. The elements of the two- and three-particle S-matrix are defined by
\[
\begin{split}
 \label{eqn:generic}
  S \ket{\Phi_{i}(a) \Phi_{j}(b)} & = \ket{\Phi_{k}(a) \Phi_{l}(b)} S_{ij}^{kl}(a,b) \; , \\
  S \ket{\Phi_{i}(a) \Phi_{j}(b) \Phi_{k}(c)} & = \ket{\Phi_{l}(a) \Phi_{m}(b) \Phi_{n}(c)} S_{ijk}^{lmn}(a,b,c) \; ,
\end{split}
\]
where $\Phi_i$ denotes the set of all fields, bosonic as well as fermionic, and $a,b,c$ denote the individual momenta. Then integrability implies
\[ \label{eqn:Smat-factorization}
 \tilde S_{ijk}^{lmn}(a,b,c)
 = \sum_{xyz}
 \tilde S_{xy}^{lm}(a,b)
 \tilde S_{iz}^{xn}(a,c)
 \tilde S_{jk}^{yz}(b,c)
 = \sum_{xyz}
 \tilde S_{yz}^{mn}(b,c)
 \tilde S_{xk}^{lz}(a,c)
 \tilde S_{ij}^{xy}(a,b)
 \; ,
\]
for the graded matrix elements $\tilde S_{ij}^{kl} = (-)^{\abs{i}\abs{j}} S_{ij}^{kl}$ and $\tilde S_{ijk}^{lmn} = (-)^{\abs{i}\abs{j} + \abs{j}\abs{k} + \abs{k}\abs{i}} S_{ijk}^{lmn}$. The intricate index flow in this equation is illustrated in \figref{fig:factorization}.
The first equality in \eqref{eqn:Smat-factorization} is the statement of the factorization of the three-particle S-matrix. This is the equation we will prove for the string sigma-model in the near-flat limit to one-loop order. The second equality in \eqref{eqn:Smat-factorization} is the Yang-Baxter equation (YBE), which is a consistency condition for the two-particle S-matrix. It is necessary but not sufficient for factorization\footnote{This is to be contrasted with the algebraic Bethe ansatz where the YBE for the universal R-matrix is sufficient for integrability. However, for the AdS string the universal R-matrix is unknown to date. For some work in this direction see \cite{Torrielli:2007mc,Heckenberger,Moriyama:2007jt}.}. The two-particle S-matrix for the string sigma-model satisfies the YBE by construction \cite{Beisert:2006qh,Arutyunov:2006yd}.

We stress that the S-matrices in \eqref{eqn:generic} and \eqref{eqn:Smat-factorization} are the total S-matrices, including all connected and disconnected parts. Factorization is a statement about the localization of the interaction, not about disconnectedness.

%%%%%%%%%%%%%%%%%%%%%%%%%%%%%%%%%%%%%%%%%%%%%%%%%%%%%%%%%%%%%%%%%%%%%%%%%%%
\paragraph{Factorized scattering and Feynman diagrams.} It is interesting to see how \eqref{eqn:Smat-factorization} comes out of a Feynman diagram computation. The first thing to note is that \eqref{eqn:Smat-factorization} implies that three-particle amplitudes vanish if the out-momenta do not pairwise equal the in-momenta. This is a consequence of the higher conserved charges. In the Feynman diagram computation, however, only energy and momentum conservation are manifest. Nevertheless, the sum of all diagrams will turn out to vanish identically except at those isolated points in phase space, where the set of out-momenta equals the set of in-momenta. At these points the amplitude diverges due to an internal propagator going on-shell. This divergence is regularized by the $i\eps$-prescription and the residue is extracted using \cite{Dorey:1996gd}
\[ \label{eqn:principle-value-formula}
  \frac{1}{\vec{p}^2 - m^2 \pm i \eps} = \mathcal{P} \frac{1}{\vec{p}^2 - m^2} \mp i \pi \delta(\vec{p}^2 - m^2) \; .
\]
The principal value terms cancel and the amplitude acquires a further $\delta$-function, which together with the $\delta$-functions from energy-momentum conservation forces each out-momen\-tum to be equal to one of the in-momenta. Though rather implicit, this is the effect of the next conserved charge.

\begin{figure}
\begin{center}
\includegraphics[scale=0.8]{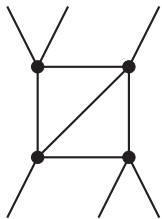}
\end{center}
\caption{\textbf{Generic two-loop diagram.} At two-loop order, there are three-to-three particle Feynman diagrams which do not split into a sequence of three two-to-two subdiagrams.}
\label{fig:generic2loop}
\end{figure}

One is tempted to further explain the factorization directly by regarding \figref{fig:factorization} as Feynman diagrams where the vertices of the three-particle process are grouped into three subdiagrams each of which corresponds to a two-particle process. The internal on-shell propagator would then be expected to be one of the lines connecting the two-particle S-matrix blobs. We will show, however, that this identification is not possible in general. This becomes obvious at higher loop orders where generic diagrams do not have the structure of the right hand side of \figref{fig:factorization} (see e.g. \figref{fig:generic2loop}).
But already at one-loop order, where this is still the case, we will see that \figref{fig:factorization} can at most be interpreted as effective diagrams after all Feynman graphs have been added together. This makes S-matrix factorization to an even more exceptional property.

%%%%%%%%%%%%%%%%%%%%%%%%%%%%%%%%%%%%%%%%%%%%%%%%%%%%%%%%%%%%%%%%%%%%%%%%%%%
%%%%%%%%%%%%%%%%%%%%%%%%%%%%%%%%%%%%%%%%%%%%%%%%%%%%%%%%%%%%%%%%%%%%%%%%%%%
\section{The near-flat-space model}
\label{sec:MS-model}

The full $\AdS_5\times\Sphere^5$ sigma-model \cite{Metsaev:1998it} is rather complicated. In light-cone gauge, which is the preferred gauge for the particle picture of the AdS/CFT correspondence, the action contains interactions of arbitrary order \cite{Frolov:2006cc}. However, in the near-flat limit introduced by Maldacena and Swanson \cite{Maldacena:2006rv}, the theory simplifies drastically but without becoming trivial. The world-sheet Lagrangian in this limit can be written as \cite{Klose:2007wq, Klose:2007rz}
\begin{eqnarray} \label{eqn:MS-action}
 \mathcal{L} & = &
  \tfrac{1}{2}(\partial Y)^2-\tfrac{m^2}{2}\,Y^2
 +\tfrac{1}{2}(\partial Z)^2-\tfrac{m^2}{2}\,Z^2
 +\tfrac{i}{2}\psi \tfrac{\partial^2+m^2}{\partial_-}\,\psi \nn \\[1mm]
 &&
 + \g\,(Y^2-Z^2)\bigbrk{(\partial _-Y)^2+(\partial _-Z)^2}
 +i\g\,(Y^2-Z^2)\psi\partial_-\psi \nn \\[1mm]
 &&
 +i\g\,\psi\bigbrk{\partial_- Y_{i'} \Gamma_{i'} + \partial_- Z_i \Gamma_i}
           \bigbrk{Y_{j'} \Gamma_{j'} - Z_j \Gamma_j}\psi \nn \\[1mm]
 &&
 -\tfrac{\g}{24}\bigbrk{\psi\Gamma_{i'j'}\psi\,\psi\Gamma_{i'j'}\psi
                      -\psi\Gamma_{ij}  \psi\,\psi\Gamma_{ij}  \psi} \; ,
\end{eqnarray}
where the coupling constant $\gamma$ is related to the 't Hooft coupling by
\be
\g = \frac{\pi}{\sqrt{\lambda}}\; .
\ee
The bosonic fields $Y_{i'=1,\ldots,4}$ and $Z_{i=5,\ldots,8}$ are the transverse excitations in $\Sphere^5$ and $\AdS_5$, respectively. The eight fermionic degrees of freedom are described by an $\grSO(8)$ Majorana-Weyl spinor $\psi$. The parameter $m$ has been introduced into \eqref{eqn:MS-action} merely to facilitate power counting and should be set to one in the results. Finally, $\partial_{\pm}$ are the usual light-cone derivatives $\partial_{\pm}=\frac{1}{2}(\partial_0\pm\partial_1)$. In \appref{app:notation} more details on the notation are given.

In essence, the near-flat limit enhances the left-moving and suppresses the right-moving modes of the string excitations by sending%
\footnote{This is the scaling of the momenta in the string normalization, which is related to the normalization of the momenta used in the spin chain picture by $p_{\mathrm{string}} = \frac{\sqrt{\lambda}}{2\pi} \, p_{\mathrm{chain}}$.}
\[ \label{eqn:near-flat-limit}
  p_- \to \infty
  \comma
  p_+ \to 0
  \quad\text{with}\quad
  p_{\pm} \lambda^{\pm 1/4} = \mbox{fixed} \; .
\]
While this decoupling of left- and right-movers leads to the significant simplifications, the near-flat-space model is nevertheless capable to interpolate between the classical giant magnon regime and the free plane-wave limit \cite{Maldacena:2006rv}. The comparison of results for the full sigma-model and the reduced model is done by scaling the plane-wave momenta in the former as in \eqref{eqn:near-flat-limit}. The quantum consistency of the near-flat-space truncation has been checked up to two-loops by computing the two-particle S-matrix \cite{Klose:2007rz}.

%%%%%%%%%%%%%%%%%%%%%%%%%%%%%%%%%%%%%%%%%%%%%%%%%%%%%%%%%%%%%%%%%%%%%%%%%%%
\paragraph{Decompactification limit.} Taking the near-flat limit of the AdS string naturally decompactifies the cylindrical world-sheet \cite{Maldacena:2006rv} such that the model \eqref{eqn:MS-action} is in fact defined on the infinite plane
\[
  \Action = \int_{-\infty}^\infty d^2\sigma\: \Lagr \; .
\]
This is no inconvenience, but just in the spirit of the world-sheet S-matrix approach to the string spectrum \cite{Klose:2006dd}. In order to sensibly define asymptotic states and study scattering amplitudes, one needs to have a non-compact world-sheet. A further requirement is the relaxation of the level-matching condition, as the S-matrix describes the scattering of single excitations which are actually unphysical string states.

%%%%%%%%%%%%%%%%%%%%%%%%%%%%%%%%%%%%%%%%%%%%%%%%%%%%%%%%%%%%%%%%%%%%%%%%%%%
\paragraph{Symmetries.} The model \eqref{eqn:MS-action} inherits the symmetries from the $\AdS_5\times\Sphere^5$ superstring in light-cone gauge \cite{Maldacena:2006rv}. With relaxed level-matching, the symmetry algebra is $\algPSU(2|2)^2 \ltimes \Reals^3$ \cite{Arutyunov:2006ak}, where $\ltimes$ symbolizes central extension. The charges of the fields under the bosonic subalgebra $\algSU(2)^4$ can be made manifest by changing to a double index notation
\[ \label{eqn:switch-notation}
  (Y_{i'},Z_{i},\psi) \longleftrightarrow (Y_{\lAA\rAA},Z_{\laa\raa},\Psi_{\lAA\raa},\Bsi_{\laa\rAA}) \; .
\]
Here $\lAA$, $\rAA$ are even and $\laa$, $\raa$ are odd $\algSU(2)$ indices, which can be combined as $A=(\lAA|\laa)$ and $\dot{A}=(\rAA|\raa)$ into the fundamental indices of the two $\algPSU(2|2)$ factors. In \appref{app:embedding} we  give the precise relationship between the two notations in \eqref{eqn:switch-notation}.

%%%%%%%%%%%%%%%%%%%%%%%%%%%%%%%%%%%%%%%%%%%%%%%%%%%%%%%%%%%%%%%%%%%%%%%%%%%
\paragraph{Two-particle S-matrix.}

The S-matrix for the AdS/CFT particle model was constructed from symmetries in \cite{Beisert:2005tm}, and written in the string basis in \cite{Arutyunov:2006yd}. The reduction to the near-flat space was investigated in \cite{Maldacena:2006rv}. Here we give the form derived in \cite{Klose:2007rz}, where it was also shown to arise from Feynman diagrams to two-loop order. It is given by the graded tensor product of two identical $\algSU(2|2)$ S-matrices
\[ \label{eqn:Smat-2p}
 \Smatrix_{A\dot{A}B\dot{B}}^{C\dot{C}D\dot{D}}(a,b) =
 (-)^{\abs{\dot{A}}\abs{B} + \abs{\dot{C}}\abs{D}} \,
 S_0(a,b) \,
 \smatrix_{AB}^{CD}(a,b) \,
 \smatrix_{\dot{A}\dot{B}}^{\dot{C}\dot{D}}(a,b) \; .
\]
The arguments $a \equiv p_{a-}$ and $b \equiv p_{b-}$ are the minus components of the particles' light-cone momenta. Up to order $\order(\g^4)$ corrections, the prefactor $S_0$ can be written as
\[ \label{eqn:Smat-pre}
  S_0(a,b) = \frac{\,\,e^{
  \frac{8i}{\pi} \g^2 \, \frac{a^3 b^3}{b^2 - a^2}
  \lrbrk{1-\frac{b^2 + a^2}{b^2 - a^2} \, \ln\frac{b}{a}}
  }}
  {1+\g^2 \, a^2 b^2 \lrbrk{\frac{b + a}{b - a}}^2} \; .
\]
The matrix part is usually parametrized as follows
\begin{align} \label{eqn:Smat-mat}
  \smatrix_{\lAA\lBB}^{\lCC\lDD} & = A \,\delta_\lAA^\lCC \delta_\lBB^\lDD
                                   + B \,\delta_\lAA^\lDD \delta_\lBB^\lCC \; , &
  \smatrix_{\lAA\lBB}^{\lcc\ldd} & = C \,\levi_{\lAA\lBB} \levi^{\lcc\ldd} \; , &
  \smatrix_{\lAA\lbb}^{\lCC\ldd} & = G \,\delta_\lAA^\lCC \delta_\lbb^\ldd \; , &
  \smatrix_{\lAA\lbb}^{\lcc\lDD} & = H \,\delta_\lAA^\lDD \delta_\lbb^\lcc \; , \\
  \smatrix_{\laa\lbb}^{\lcc\ldd} & = D \,\delta_\laa^\lcc \delta_\lbb^\ldd
                                   + E \,\delta_\laa^\ldd \delta_\lbb^\lcc \; , &
  \smatrix_{\laa\lbb}^{\lCC\lDD} & = F \,\levi_{\laa\lbb} \levi^{\lCC\lDD} \; , &
  \smatrix_{\laa\lBB}^{\lcc\lDD} & = L \,\delta_\laa^\lcc \delta_\lBB^\lDD \; , &
  \smatrix_{\laa\lBB}^{\lCC\ldd} & = K \,\delta_\laa^\ldd \delta_\lBB^\lCC \; , \nn
\end{align}
where the exact coefficient functions are given by
\begin{align} \label{eqn:Smat-coeffs}
  A(a,b) & =  1 + i\g \, a\,b\,\frac{b-a}{b+a} \; , &
  B(a,b) & = -E(a,b) = 4i\g\, \frac{a^2\,b^2}{b^2-a^2} \; , \nn \\
  D(a,b) & =  1 - i\g \, a\,b\,\frac{b-a}{b+a} \; , &
  C(a,b) & =  F(a,b) = 2i\g\, \frac{a^{3/2}\,b^{3/2}}{b+a} \; , \\
  G(a,b) & =  1 + i\g \, a\,b \; , &
  H(a,b) & =  K(a,b) = 2i\g\, \frac{a^{3/2}\,b^{3/2}}{b-a} \; , \nn \\[3mm]
  L(a,b) & =  1 - i\g \, a\,b \; . \nn
\end{align}

%%%%%%%%%%%%%%%%%%%%%%%%%%%%%%%%%%%%%%%%%%%%%%%%%%%%%%%%%%%%%%%%%%%%%%%%%%%
%%%%%%%%%%%%%%%%%%%%%%%%%%%%%%%%%%%%%%%%%%%%%%%%%%%%%%%%%%%%%%%%%%%%%%%%%%%
\section{Symmetries of the three-particle S-matrix}
\label{sec:symmetries}

The three-particle S-matrix turns out to be highly restricted due to the $\algPSU(2|2)^2\ltimes\Reals^3$ symmetry of the model.
The representation theory of this algebra has been worked out in \cite{Beisert:2006qh}, where it was found that the triple tensor product of the fundamental representation $\rep{4}$ of $\algPSU(2|2)\ltimes\Reals^3$ decomposes into two 32-dimensional irreducible representations
\[
  \raisebox{-0.9mm}{\includegraphics[scale=0.7]{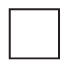}} \;\otimes\;
  \raisebox{-0.9mm}{\includegraphics[scale=0.7]{yang-00}} \;\otimes\;
  \raisebox{-0.9mm}{\includegraphics[scale=0.7]{yang-00}} \;=\;
  \raisebox{-2.8mm}{\includegraphics[scale=0.7]{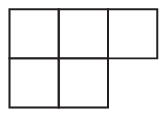}} \;\oplus\;
  \raisebox{-4.6mm}{\includegraphics[scale=0.7]{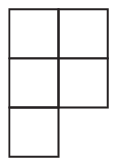}} \; .
\]
Taking also the other $\algPSU(2|2)$ factor into account we conclude that the three-particle S-matrix is a sum of four projectors
\[ \label{eqn:S-matrix-proj}
  \Smatrix
  = C_1 \,
  P_{\lrbrk{\raisebox{-0.1mm}{\includegraphics[scale=0.2]{yang-10}},
            \raisebox{-0.1mm}{\includegraphics[scale=0.2]{yang-10}}}}
  + C_2 \,
  P_{\lrbrk{\raisebox{-0.1mm}{\includegraphics[scale=0.2]{yang-10}},
            \raisebox{-0.7mm}{\includegraphics[scale=0.2]{yang-01}}}}
  + C_3 \,
  P_{\lrbrk{\raisebox{-0.7mm}{\includegraphics[scale=0.2]{yang-01}},
            \raisebox{-0.1mm}{\includegraphics[scale=0.2]{yang-10}}}}
  + C_4 \,
  P_{\lrbrk{\raisebox{-0.7mm}{\includegraphics[scale=0.2]{yang-01}},
            \raisebox{-0.7mm}{\includegraphics[scale=0.2]{yang-01}}}}
  \; .
\]
The coefficients $C_i$ are functions of the three in-particle momenta which are measured by the projectors $P_i$ from the state they act upon. In order to determine these coefficients, and hence fix the entire three-particle S-matrix, it is most convenient to consider the scattering of eigentstates of the projectors $P_i$, for instance
\begin{align}
\label{eqn:hw-states}
  & \Smatrix  \,\ket{    Y_{1\dot{1}}(a) \,    Y_{1\dot{1}}(b) \,    Y_{1\dot{1}}(c) }
  = C_1(a,b,c)\,\ket{    Y_{1\dot{1}}(a) \,    Y_{1\dot{1}}(b) \,    Y_{1\dot{1}}(c) } \; , \\
  & \Smatrix  \,\ket{ \Psi_{1\dot{3}}(a) \, \Psi_{1\dot{3}}(b) \, \Psi_{1\dot{3}}(c) }
  = C_2(a,b,c)\,\ket{ \Psi_{1\dot{3}}(a) \, \Psi_{1\dot{3}}(b) \, \Psi_{1\dot{3}}(c) } \; , \nn \\
  & \Smatrix  \,\ket{ \Bsi_{3\dot{1}}(a) \, \Bsi_{3\dot{1}}(b) \, \Bsi_{3\dot{1}}(c) }
  = C_3(a,b,c)\,\ket{ \Bsi_{3\dot{1}}(a) \, \Bsi_{3\dot{1}}(b) \, \Bsi_{3\dot{1}}(c) } \; , \nn \\
  & \Smatrix  \,\ket{    Z_{3\dot{3}}(a) \,    Z_{3\dot{3}}(b) \,    Z_{3\dot{3}}(c) }
  = C_4(a,b,c)\,\ket{    Z_{3\dot{3}}(a) \,    Z_{3\dot{3}}(b) \,    Z_{3\dot{3}}(c) } \; . \nn
\end{align}
The states in \eqref{eqn:hw-states} are just the highest weight states of the respective modules.

We note that the Lagrangian is not invariant under the exchange of the two $\algPSU(2|2)$ factors: some interactions are symmetric while others are anti-symmetric. Therefore we cannot further reduce the expansion \eqref{eqn:S-matrix-proj} by setting%
\footnote{Nevertheless, we will find $C_2=C_3$ as a result of the one-loop computation in \secref{sec:3p-scatt}. However, we may not use this as an input here.}
$C_2$ equal to $C_3$. The lack of this symmetry can also be seen from the two-particle S-matrix, which sometimes acquires a sign when undotted and dotted indices are interchanged:
\[ \label{eqn:Smat-non-symmetry}
 \Smatrix_{\dot{A}A\dot{B}B}^{\dot{C}C\dot{D}D} =
 (-)^{\abs{\dot{A}}\abs{B} + \abs{A}\abs{\dot{B}} + \abs{\dot{C}}\abs{D} + \abs{C}\abs{\dot{D}}} \,
 \Smatrix_{A\dot{A}B\dot{B}}^{C\dot{C}D\dot{D}} \; .
\]

The S-matrix elements $C_i$ can be extracted from
\[ \label{eqn:C1}
C_1(a,b,c) \sum_{\sigma(d,e,f)} \delta_{ad} \, \delta_{be} \, \delta_{cf}
=
\bra{Y_{1\dot{1}}(f) \, Y_{1\dot{1}}(e) \, Y_{1\dot{1}}(d)}
\, \Smatrix \,
\ket{Y_{1\dot{1}}(a) \, Y_{1\dot{1}}(b) \, Y_{1\dot{1}}(c)} \; ,
\]
etc. The sum runs over all the permutations of $(d,e,f)$ and $\delta_{ab}\equiv 2\pi\delta(a-b)$.

For practical reasons it is convenient to perform the actual computation for $\algSO(4)^2$ fields, in terms of which the amplitude \eqref{eqn:C1} reads
\be
\label{eqn:fourampl}
\bra{Y_{1\dot{1}}\,Y_{1\dot{1}}\,Y_{1\dot{1}}}
\,\Smatrix \,
\ket{Y_{1\dot{1}}\,Y_{1\dot{1}}\,Y_{1\dot{1}}} \eq
 \bra{Y_1\,Y_1\,Y_1}\,\Smatrix\,\ket{Y_1\,Y_1\,Y_1}
-\bra{Y_1\,Y_1\,Y_1}\,\Smatrix\,\ket{Y_1\,Y_4\,Y_4} \nl
-\bra{Y_1\,Y_1\,Y_1}\,\Smatrix\,\ket{Y_4\,Y_1\,Y_4}
-\bra{Y_1\,Y_1\,Y_1}\,\Smatrix\,\ket{Y_4\,Y_4\,Y_1} \; ,
\ee
where the momentum arguments are as in \eqref{eqn:C1}. In \secref{sec:3p-scatt} we present some details of the computation of these expectation values and summarize the results for all processes \eqref{eqn:hw-states}. In \secref{sec:factorization} we show that these matrix elements factorize into products of two-particle S-matrices. On the basis of the $\algPSU(2|2)^2\ltimes\Reals^3$ symmetry, this proves the factorization of the full three-particle S-matrix.

We should stress that we are not going to write down the full functions $C_i$, but only those parts which describe processes where all particles are involved in the interaction. In the language of Feynman diagrams these are the terms corresponding to \emph{fully connected graphs}. Being interested in the factorization of the S-matrix, these are the crucial terms. Contributions from processes where one particle is merely a spectator factorize trivially. They correspond to disconnected Feynman diagrams and probe only the two-particle S-matrix. We refrain from computing these contributions, since it has already been explicitly verified in \cite{Klose:2007rz} that the two-particle S-matrix arises from Feynman diagrams up to order $\g^3$. In \secref{sec:factorization}, we explain how to separate connected and disconnected contributions to the S-matrix \eqref{eqn:S-matrix-proj} from each other.

In fact, there is no real gain in knowing the actual expressions for $C_i$ as long as one does not have the explicit form of the projectors $P_i$. We are making use of the S-matrix representation \eqref{eqn:S-matrix-proj} only to prove factorization. After having done so, it is much more convenient to utilize factorization and compute three-particle amplitudes from the product of two-particle S-matrices \eqref{eqn:Smat-factorization}.

%%%%%%%%%%%%%%%%%%%%%%%%%%%%%%%%%%%%%%%%%%%%%%%%%%%%%%%%%%%%%%%%%%%%%%%%%%%
%%%%%%%%%%%%%%%%%%%%%%%%%%%%%%%%%%%%%%%%%%%%%%%%%%%%%%%%%%%%%%%%%%%%%%%%%%%
\newpage
\section{Three-particle scattering}
\label{sec:3p-scatt}

In this section, we compute the three-particle scattering amplitudes from Feynman diagrams. In order to prove factorization, it is sufficient to calculate the ``highest weight processes'' \eqref{eqn:hw-states}. The results are summarized in \secref{sec:essential-processes}, and some further processes are discussed in \secref{sec:bonus-processes}.

For describing the computation we take, as an illustrative example, the process
\[ \label{eqn:111-111}
  Y_1(a) Y_1(b) Y_1(c) \rightarrow Y_1(d) Y_1(e) Y_1(f) \; ,
\]
which is the first part in \eqref{eqn:fourampl}. We denote the amplitude as
\[ \label{eqn:defamp}
  \Amp(\pin) \equiv \Amp(a,b,c,d,e,f) = \braket{Y_1(f) Y_1(e) Y_1(d)}{Y_1(a) Y_1(b) Y_1(c)}_{\mathrm{connected}} \; .
\]
As discussed at the end of \secref{sec:symmetries}, we can restrict ourselves to the connected part of the amplitude, because the disconnected part factorizes trivially. The in-coming momenta are ordered as $c>b>a>0$. This corresponds to $p_a > p_b > p_c$, so that the state $\ket{Y_1(a) Y_1(b) Y_1(c)}$ where the fields are arranged as $x_a < x_b < x_c$ is indeed an initial state. The out-going momenta $d,e,f>0$ are constrained by momentum and energy conservation
\[ \label{eqn:emc}
 a + b + c = d + e + f
 \comma
 \frac{1}{a} + \frac{1}{b} + \frac{1}{c} = \frac{1}{d} + \frac{1}{e} + \frac{1}{f} \; ,
\]
where the second equation receives corrections at two loops \cite{Klose:2007rz}. These equations force the out-momenta to lie on a planar curve, cf. \figref{fig:phasespace} in \appref{app:phasespace} where the phase space is discussed in detail.

The key point of factorized scattering is that the amplitudes will be of the form
\[ \label{eqn:exp-amp}
  \Amp(\pin) = \sum_{\sigma(d,e,f)} \Samp_\sigma(a,b,c)\; \delta_{ad}\, \delta_{be} \,\delta_{cf} \, ,
\]
i.e. they vanish identically on most of the phase space and peak at those particular points where the set of the in-coming momenta $\{a,b,c\}$ is equal to the set of the out-going momenta $\{d,e,f\}$. This form is a manifestation of the underlying higher conservation laws. The functions $\Samp_\sigma(a,b,c)$ are by definition the three-particle S-matrix elements, where the permutation $\sigma$ determines the ordering of the flavor indices in the out-state (keeping the momenta always in the order $a,b,c$). This means that a single amplitude determines up to six S-matrix elements.

By showing that the $3 \rightarrow 3$ amplitude is of the form \eqref{eqn:exp-amp}, one shows at the same time that there is no $2 \rightarrow 4$ particle production. The amplitude for the latter can be deduced from the former by analytically continuing one of the in-coming momenta to negative values. In this case, however, the condition $\{a,b,c\} = \{d,e,f\}$ cannot be satisfied, because the momenta in the first set are all positive while one momentum in the second set is negative. Hence, the particle production amplitude vanishes identically.

%%%%%%%%%%%%%%%%%%%%%%%%%%%%%%%%%%%%%%%%%%%%%%%%%%%%%%%%%%%%%%%%%%%%%%%%%%%
\subsection{Tree-level}

\begin{figure}
\begin{center}
\includegraphics[scale=0.8]{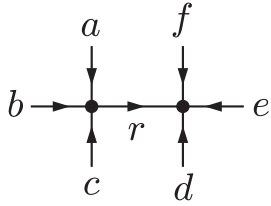}
\end{center}
\caption{\textbf{Tree-level diagram.} The label $r$ counts the derivatives acting onto the internal propagator.}
\label{fig:tree}
\end{figure}

At tree-level the amplitude is computed from diagrams of the kind drawn in \figref{fig:tree}. For the process \eqref{eqn:defamp} we find
\[ \label{eqn:tree-amplitude}
  \Amp^{\mathrm{tree}}(\pin) = -i\g^2 \, \frac{1}{\sqrt{64abcdef}} \, \frac{1}{2!\,3!^2} \sum_{\sigma(\pou)} F(\pou) \, \prop_0(\pou) \; ,
\]where
\[
  F(\pin) = 16 \, \bigbrk{ a^2 + b^2 + c^2 + a b + b c + c a } \, \bigbrk{ d^2 + e^2 + f^2 + de + ef + fd } \; .
\]
and $\prop_0$ is the tree-diagram (cf.~\eqref{eqn:tree})
\[ \label{eqn:prop0}
 \prop_0(\pin) = \frac{\delta^2(\vecpin)}{(\vec{a}+\vec{b}+\vec{c})^2 - m^2 + i \eps} \; .
\]
The sum in \eqref{eqn:tree-amplitude} is taken over all permutations of $\pou \equiv (a,b,c,-d,-e,-f)$. Since the summand is symmetric in $(a,b,c)$ and in $(d,e,f)$ and under the exchange $(a,b,c)\leftrightarrow(d,e,f)$, one can restrict the sum to permutations under which the summand is not symmetric (there are 10 such permutations) and drop the factor $\frac{1}{2!\,3!^2}$. The first fraction in \eqref{eqn:tree-amplitude} originates from the wave-function normalization of the external particles.

After performing the sum in \eqref{eqn:tree-amplitude}, one can use energy-momentum conservation to show that the amplitude indeed vanishes if the sets of in- and out-momenta are different:
\[
 \Amp^{\mathrm{tree}}(\pin) = 0 \quad\mbox{for}\quad \{a,b,c\} \not= \{d,e,f\} \; .
\]
At the points where $\{a,b,c\} = \{d,e,f\}$ the amplitude becomes divergent when $\eps\to0$ in \eqref{eqn:prop0}. The divergences originate from terms where the momentum of the internal propagator $\prop_0$ is equal to one of the external momenta and therefore goes on-shell. The divergence is of $\delta$-function-type and its residue can be extracted by means of the principal value formula \eqref{eqn:principle-value-formula}. In the sum \eqref{eqn:tree-amplitude} the principal value terms cancel because of energy-momentum conservation and we are left with an additional $\delta(\vec{p}^2-m^2)$-function which sets the internal momentum $\vec{p}$ of the corresponding diagram on-shell. The factorized form \eqref{eqn:exp-amp} arises from combining this $\delta$-function with the overall energy-momentum conservation $\delta^{(2)}(\vecpin)$ contained in \eqref{eqn:prop0}. For the case at hand we obtain
\[ \label{eqn:tree-amplitude-result}
  \Amp^{\mathrm{tree}}(\pin) = - \frac{\g^2}{4 m^4} \, \frac{a b c \, G(a,b,c)}{(a+c)(c-b)(b-a)}
                                 \sum_{\sigma(d,e,f)} \delta_{ad} \, \delta_{be} \, \delta_{cf}
\]
with
\[
  G(a,b,c) = 16 \bigsbrk{ 2 a b c (a-b+c) + a^3 (b-c) + b^3 (a+c) + c^3 (b-a) } \; .
\]
This result is a special case of \eqref{eqn:exp-amp}, where the coefficients $\Samp_\sigma$ are actually independent of the permutation $\sigma$ which is due to the fact that all involved fields are of the same flavor.

It is interesting to compare the result \eqref{eqn:tree-amplitude-result} to the corresponding result in $\phi^4$-theory which is reviewed in the context of integrability in \cite{Dorey:1996gd}. There, the amplitude is given by \eqref{eqn:tree-amplitude} with $F(\pin) \equiv 1$. As a consequence, $\Amp^{\mathrm{tree}}(\pin)$ is not of factorized form. It is given by \eqref{eqn:tree-amplitude-result} setting $G(a,b,c) \equiv 1$ but with an additional constant piece. This constant can, however, be canceled from the amplitude by adding a six-point vertex to the action. Continuing this process to the scattering of more and more particles, turns $\phi^4$-theory into sinh-Gordon. Here, in the near-flat-space model, the amplitude vanishes without higher order interactions.

%%%%%%%%%%%%%%%%%%%%%%%%%%%%%%%%%%%%%%%%%%%%%%%%%%%%%%%%%%%%%%%%%%%%%%%%%%%
\subsection{One-loop}

\begin{figure}
\begin{center}
\subfigure[bubble]{\includegraphics[scale=0.8]{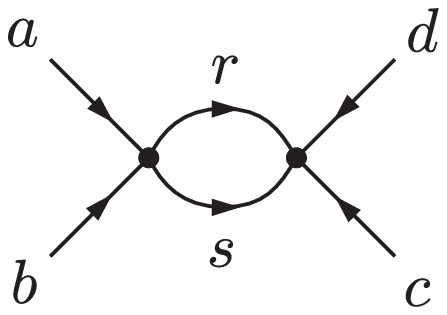}\label{fig:bubble}}\qquad
\subfigure[dog]{   \includegraphics[scale=0.8]{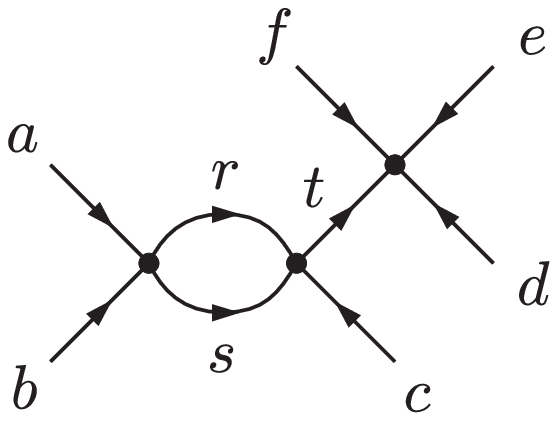}   \label{fig:dog}}   \qquad
\subfigure[sun]{   \includegraphics[scale=0.8]{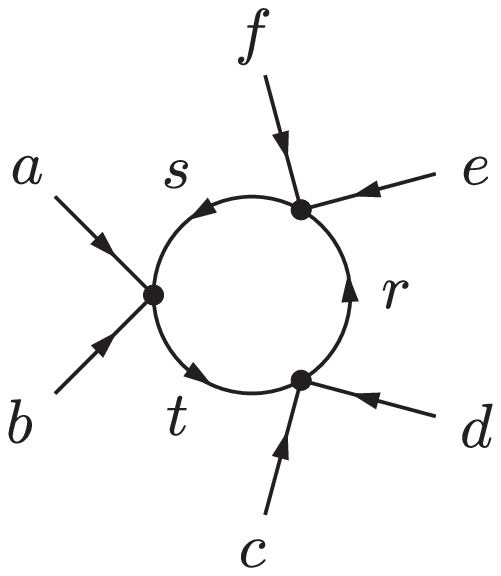}   \label{fig:sun}}
\end{center}
\caption{\textbf{One-loop diagrams.} The labels $r$, $s$ and $t$ count the derivatives acting onto the corresponding internal propagator. The corresponding integrals are evaluated in \protect\appref{app:diagrams}.}
\label{fig:one-loop}
\end{figure}

The one-loop amplitude is given by two sets of diagrams, the ``dogs'' (\figref{fig:dog}) and the ``suns'' (\figref{fig:sun}),
\[ \label{eqn:1loop-dog-sun}
  \Amp^{\mathrm{1-loop}}(\pin) = \Amp^{\mathrm{dog}}(\pin) + \Amp^{\mathrm{sun}}(\pin) \; .
\]
As before, we describe the general computation and give explicit results for the sample process \eqref{eqn:111-111}.

%%%%%%%%%%%%%%%%%%%%%%%%%%%%%%%%%%%%%%%%%%%%%%%%%%%%%%%%%%%%%%%%%%%%%%%%%%%
\paragraph{The dogs.} The combinatorics gives a linear combination of $\dog_{rst}$ with up to five derivatives ($r+s+t\le5$). The result can be simplified by making use of the identities \eqref{eqn:dog-relation} which relate $\dog_{rst}$'s with different numbers of derivatives. In all considered cases it was possible to reduce the expressions such that only $\dog_{000}$ remains. For the amplitude of \eqref{eqn:111-111} the result is
\[
\begin{split}
\Amp^{\mathrm{dog}}(\pin) =
\frac{1}{\sqrt{64abcdef}} \, \frac{32 \g^3}{2!\,3!} \, \sum_{\sigma(\pou)} &
\bigbrk{(a + b)^2 (a^2 + a b + b^2) + (a + b + c) c (a^2 - 6 a b + b^2)} \\[-4mm]
& \quad \times \bigbrk{ d^2 + e^2 + f^2 + de + ef + fd } \dog_{000}(\pou) \; ,
\end{split}
\]
where the sum is taken over all permutations of the momenta $\{a,b,c,-d,-e,-f\}$. As in the tree-level case, the prefactor $\frac{1}{2!\,3!}$ can be removed, if the sum is restricted to permutations under which the summand in not symmetric. Here the symmetries are permutation of $(a,b)$ and of $(d,e,f)$.

%%%%%%%%%%%%%%%%%%%%%%%%%%%%%%%%%%%%%%%%%%%%%%%%%%%%%%%%%%%%%%%%%%%%%%%%%%%
\paragraph{The suns.} Generically one finds very lengthy expressions in $\sun_{rst}$ with up to six derivatives. Using the identities \eqref{eqn:sun-relation}, it turned out that all considered cases allowed a reduction of the result to $\sun_{rst}$'s with at most two derivatives. For the process \eqref{eqn:111-111} we are left with
\[
\Amp^{\mathrm{sun}}(\pin) = \frac{1}{\sqrt{64abcdef}} \, \frac{16 \g^3}{2!^3} \sum_{\sigma(\pou)}
\Bigsbrk{ V(\pou) \sun_{000}(\pou) - U(\pou) \sun_{200}(\pou) } \; ,
\]
where
\be
U(\pin) \eq
(a + b)(c + d)(e + f)^2
+ (a b + c d)(e^2 + 8e f + f^2) \nl
+ (a^2 + b^2 + c^2 + d^2)(e^2 + e f + f^2) \; ,
\\
V(\pin)\eq
6abcdef+2e^2f^2(a+b)(c+d)+(e^3+f^3)\big[ab(c+d)+cd(a+b)\big]\nl
+(e^2+f^2)\big[(a^2+b^2)(c^2+d^2)+2ab(c+d)^2+2cd(a+b)^2-abcd\nl
+(a+b)(c+d)^3+(a+b)^3(c+d)+ef(a+b)(c+d)\big] \; .
\ee
While it is not hard to evaluate the sun diagrams $\sun_{rst}$ for any number of derivatives, the cases with $r+s+t\le2$ are very special. In these cases the simplest form of a cutting rule \eqref{eqn:cutting-rule} \cite{Kallen:1965} applies, which makes it is possible to write the sun $\sun_{rst}$ as a linear combination of tree-level diagrams $\prop_r$ multiplied by bubbles $\bubble_{rs}$ (see \figref{fig:bubble}).

%%%%%%%%%%%%%%%%%%%%%%%%%%%%%%%%%%%%%%%%%%%%%%%%%%%%%%%%%%%%%%%%%%%%%%%%%%%
\paragraph{One-loop result.} The results for the integrals $\dog_{rst}(\pin)$ and $\sun_{rst}(\pin)$ are given in \appref{app:diagrams}. Both the dog and the sun can be written as products of a bubble and a propagator. For the dog this is obvious by looking at the diagram, \figref{fig:dog}, while for the sun this happens after applying the cutting rule \eqref{eqn:cutting-rule} mentioned above. A peculiarity of this cutting procedure is that the $i\eps$-prescription for the propagator becomes momentum dependent, see \eqref{eqn:momentum-dependent-eps}.

All bubbles are finite, but -- exactly as at the tree-level -- the propagators become divergent when its momentum goes on-shell. The poles can again be extracted using the principal value formula \eqref{eqn:principle-value-formula}. The partial one-loop amplitudes for \eqref{eqn:111-111} are
\begin{align}
\label{eqn:oneloop-sample-dog}
\Amp^{\mathrm{dog}}(\pin) & = + \Ramp(\pin)
-\!\!\sum_{\sigma(a,b,c)} \frac{4 i \g^3 a^3 b^2 c}{\lrabs{a^2-c^2}}\; \frac{a^2+b^2}{a^2-b^2}\;
\biggsbrk{ \frac{a^4 + 2a^3b + 10a^2b^2 +2ab^3 + b^4}{a^2-b^2} \\
& \hspace{30mm} - \frac{4i}{\pi} \frac{ab}{\lrabs{a^2-b^2}}\lrbrk{a^2-b^2 + (a^2+b^2)\ln\frac{b}{a}} }
  \times\!\!\sum_{\sigma(d,e,f)}\delta_{ad}\delta_{be}\delta_{cf} \; , \nn % \\[2mm]
\end{align}
\begin{align}
\label{eqn:oneloop-sample-sun}
\Amp^{\mathrm{sun}}(\pin) & = - \Ramp(\pin)
-\frac{8i\g^3a^2b^2c^2}{(a^2-b^2)(b^2-c^2)(c^2-a^2)} \\[2mm]
& \hspace{20mm} \times \biggsbrk{\, 24a^2b^2c^2 + a^4 b^2 + a^2 b^4 + a^4 c^2 + a^2 c^4 + b^4 c^2 + c^2 b^4 \nn \\
& \hspace{25mm} - a^3 b^2 c + a^2 b^3 c + a^3 b c^2 + a b^3 c^2 + a^2 b c^3 - a b^2 c^3 }
  \times\!\!\sum_{\sigma(d,e,f)}\delta_{ad}\delta_{be}\delta_{cf} \; . \nn
\end{align}
In these expressions, $\Ramp(\pin) = R^{\mathrm{dog}}(\pin) \, \delta^{(2)}(\vecpin) = R^{\mathrm{sun}}(\pin) \, \delta^{(2)}(\vecpin)$ is a function with support on the phase space which drops out in the final one-loop amplitude \eqref{eqn:1loop-dog-sun}. $R^{\mathrm{dog}}(\pin)$ and $R^{\mathrm{sun}}(\pin)$ are rational functions multiplied by logarithms of various ratios of the momenta. They are non-singular for $a \not= b \not= c \not= a$ and the cancelation happens upon energy-momentum conservation between terms with the same momentum flowing through the bubble.

%%%%%%%%%%%%%%%%%%%%%%%%%%%%%%%%%%%%%%%%%%%%%%%%%%%%%%%%%%%%%%%%%%%%%%%%%%%
\subsection{Summary for highest weight processes}
\label{sec:essential-processes}

We write the tree-level and one-loop results for all amplitudes \eqref{eqn:hw-states} that are needed to prove the factorization.

%%%%%%%%%%%%%%%%%%%%%%%%%%%%%%%%%%%%%%%%%%%%%%%%%%%%%%%%%%%%%%%%%%%%%%%%%%%
\subsubsection*{$Y_{1\dot{1}}(a) Y_{1\dot{1}}(b) Y_{1\dot{1}}(c)
     \rightarrow Y_{1\dot{1}}(d) Y_{1\dot{1}}(e) Y_{1\dot{1}}(f)$}

The other three contributions \eqref{eqn:fourampl} to this amplitude are computed precisely as in the example above. The total connected amplitudes are found to be
\begin{align}
  \Amp^{\mathrm{tree}}(\pin) & =
  -4\,\g^2\,a\,b\,c
\biggsbrk{
a\frac{(a+b)(a+c)}{(a-b)(a-c)} +
b\frac{(a+b)(b+c)}{(a-b)(b-c)} +
c\frac{(a+c)(b+c)}{(a-c)(b-c)}
} \\
& \hspace{15mm} \times \sum_{\sigma(d,e,f)}\delta_{ad}\delta_{be}\delta_{cf} \; , \nn \\
\Amp^{\mathrm{dog}}(\pin) & = + \Ramp(\pin)
-\!\!\sum_{\sigma(a,b,c)} \frac{4i\g^3 a^3 b^2 c}{a^2-b^2}
\lrabs{\frac{a+c}{a-c}}
\biggsbrk{
\frac{(a+b)^3}{a-b} \\
&\hspace{30mm} -\frac{4i}{\pi} \frac{ab}{\lrabs{a^2-b^2}}
\lrbrk{a^2-b^2+\lrbrk{a^2+b^2}\ln\frac{b}{a}}
} \times \sum_{\sigma(d,e,f)}\delta_{ad}\delta_{be}\delta_{cf} \; , \nn \\
\Amp^{\mathrm{sun}}(\pin) & = - \Ramp(\pin)
+8i\,\g^3\,a^2 b^2 c^2 \, \frac{(a+b)(a+c)(b+c)}{(a-b)(a-c)(b-c)}
\times \sum_{\sigma(d,e,f)}\delta_{ad}\delta_{be}\delta_{cf} \; .
\end{align}

%%%%%%%%%%%%%%%%%%%%%%%%%%%%%%%%%%%%%%%%%%%%%%%%%%%%%%%%%%%%%%%%%%%%%%%%%%%
\subsubsection*{$Z_{3\dot{3}}(a) Z_{3\dot{3}}(b) Z_{3\dot{3}}(c)
     \rightarrow Z_{3\dot{3}}(d) Z_{3\dot{3}}(e) Z_{3\dot{3}}(f)$}

The amplitude for this process is given by the previous one after changing the overall sign of the one-loop results.
This is an immediate consequence of the anti-symmetry of the interaction Lagrangian \eqref{eqn:MS-action} under the exchange of the $Y$- and $Z$-bosons.

%%%%%%%%%%%%%%%%%%%%%%%%%%%%%%%%%%%%%%%%%%%%%%%%%%%%%%%%%%%%%%%%%%%%%%%%%%%
\subsubsection*{$\Psi_{1\dot{3}}(a) \Psi_{1\dot{3}}(b) \Psi_{1\dot{3}}(c)
     \rightarrow \Psi_{1\dot{3}}(d) \Psi_{1\dot{3}}(e) \Psi_{1\dot{3}}(f)$}

The explicit computation in the fashion of the example above produces the unspectacular result
\[ \label{eqn:33-feynman}
  \Amp^{\mathrm{tree}}(\pin) = 0
  \comma
  \Amp^{\mathrm{dog}}(\pin) = 0
  \comma
  \Amp^{\mathrm{sun}}(\pin) = 0 \; .
\]
It can be explained as follows. At tree-level, the only contribution can come from using two four-fermion vertices. Spelling out this vertex by using the explicit Dirac-matrices \eqref{eqn:Diracmat} as
\[
  8\g \brk{ \psi_1 \psi_2 \psi_3 \psi_4 - \psi_5 \psi_6 \psi_7 \psi_8 }
\]
and converting the external particles to $\algSO(8)$ notation by means of \eqref{eqn:su2-ferm-ket}, shows that the legs cannot be connected in a non-vanishing way. The dog amplitude also requires the use of at least one four-fermion vertex and is zero for the same reason. Nevertheless, there are non-trivial sun diagrams. They are built from mixed vertices, where each vertex is connected to one in-coming and one out-going external line, and where all the internal propagators are bosonic. These diagrams, however, sum to zero because of the aforementioned $Y\leftrightarrow Z$ anti-symmetry of the interactions.

Note that \eqref{eqn:33-feynman} does not imply that the quantum corrections to $C_2$ in \eqref{eqn:S-matrix-proj} vanish, rather there will be contributions from disconnected diagrams at order $\g^2$. These are given by the two-particle S-matrix elements computed in \cite{Klose:2007rz}.

%%%%%%%%%%%%%%%%%%%%%%%%%%%%%%%%%%%%%%%%%%%%%%%%%%%%%%%%%%%%%%%%%%%%%%%%%%%
\subsubsection*{$\Bsi_{3\dot{1}}(a) \Bsi_{3\dot{1}}(b) \Bsi_{3\dot{1}}(c)
     \rightarrow \Bsi_{3\dot{1}}(d) \Bsi_{3\dot{1}}(e) \Bsi_{3\dot{1}}(f)$}

For the same reason as above we find also here
\[
  \Amp^{\mathrm{tree}}(\pin) = 0
  \comma
  \Amp^{\mathrm{dog}}(\pin) = 0
  \comma
  \Amp^{\mathrm{sun}}(\pin) = 0 \; .
\]
Using that also the two-particle S-matrix elements for the processes $\Psi_{1\dot{3}}(a) \Psi_{1\dot{3}}(b) \rightarrow \Psi_{1\dot{3}}(c) \Psi_{1\dot{3}}(d)$ and $\Bsi_{3\dot{1}}(a) \Bsi_{3\dot{1}}(b) \rightarrow \Bsi_{3\dot{1}}(c) \Bsi_{3\dot{1}}(d)$ agree \cite{Klose:2007rz}, we find that $C_2=C_3$ in \eqref{eqn:S-matrix-proj} at least to order $\g^3$.

%%%%%%%%%%%%%%%%%%%%%%%%%%%%%%%%%%%%%%%%%%%%%%%%%%%%%%%%%%%%%%%%%%%%%%%%%%%
\subsection{Summary for mixed processes}
\label{sec:bonus-processes}

The highest weight processes computed in the previous subsection are distinguished from generic ones by their simplicity. One might worry that the factorization of the highest weight amplitudes is therefore very special and depends on cancelations that do not occur in general. Even though the power of the centrally extended supersymmetry algebra takes care of the factorization of all other three-particle S-matrix elements, it is nevertheless reassuring to verify explicitly the factorization of some asymmetric and more complicated cases. In this subsection we discuss an annihilation-type process
\[ \label{eqn:mixed-process}
  Y_1 \, \psi_1 \, \psi_6 \longleftrightarrow Z_7 \, Y_2 \, Y_3 \; ,
\]
that involves particles from all $\algSU(2)$ sectors.

%%%%%%%%%%%%%%%%%%%%%%%%%%%%%%%%%%%%%%%%%%%%%%%%%%%%%%%%%%%%%%%%%%%%%%%%%%%
\subsubsection*{$Y_1(a) \psi_1(b) \psi_6(c) \rightarrow Z_7(d) Y_2(e) Y_3(f)$}

The computation of this amplitude proceeds just as before. The result is given by
\be \label{eqn:mixed-forward}
\Amp^{\mathrm{tree}}(\pin) \eq
  - \frac{\g^2 a^{3}b^{3/2}c^{3/2}}{(b-a)(c-a)} \, \bigsbrk{
       \delta_{ad} \delta_{be} \delta_{cf}
     - \delta_{ad} \delta_{bf} \delta_{ce}
     + \delta_{ae} \delta_{bd} \delta_{cf}
     - \delta_{af} \delta_{bd} \delta_{ce} } \nl
  + \frac{\g^2 a^{3}b^{3/2}c^{3/2}}{(a+b)(a+c)} \, \bigsbrk{
       \delta_{ae} \delta_{bf} \delta_{cd}
     - \delta_{af} \delta_{be} \delta_{cd} } \; , % \\[3mm]
\ee
\be
\Amp^{\mathrm{1-loop}}(\pin) \eq
  + \frac{i\g^3a^{4}b^{3/2}c^{3/2}\bigsbrk{c(b-c)+a(b+c)}}{(b-a)(c^2-a^2)} \,
    \bigsbrk{ \delta_{ad} \delta_{be} \delta_{cf}
            - \delta_{ad} \delta_{bf} \delta_{ce}} \nl
  - \frac{i\g^3a^4b^{3/2}c^{3/2}\bigsbrk{a(b-c)+c(b+c)}}{(b-a)(c^2-a^2)} \,
    \bigsbrk{ \delta_{ae} \delta_{bd} \delta_{cf}
            - \delta_{af} \delta_{bd} \delta_{ce}} \nl
  + \frac{i\g^3a^4b^{3/2}c^{3/2}\bigsbrk{a(b-c)+b(b+c)}}{(b^2-a^2)(a+c)} \,
    \bigsbrk{ \delta_{ae} \delta_{bf} \delta_{cd}
            - \delta_{af} \delta_{be} \delta_{cd}} \; .
\ee
We note that the one-loop amplitude is the sum of two non-trivial contributions from $\Amp^{\mathrm{dog}}$ and $\Amp^{\mathrm{sun}}$ which we report in \appref{app:scatt-amp}. In contrast to the highest weight cases, this amplitude gives information about six different S-matrix elements. We show in the next section that all of them factorize into products of two-particle S-matrix elements.

%%%%%%%%%%%%%%%%%%%%%%%%%%%%%%%%%%%%%%%%%%%%%%%%%%%%%%%%%%%%%%%%%%%%%%%%%%%
\subsubsection*{$Z_7(a) Y_2(b) Y_3(c) \rightarrow Y_1(d) \psi_1(e) \psi_6(f)$}

The amplitude for the reversed process is found from the same Wick contractions as for the original one. The S-matrix elements, however, turn out to be quite different:
\be \label{eqn:mixed-reversed}
\Amp^{\mathrm{tree}}(\pin) \eq
  - \frac{\g^2 a^{3}b^{3/2}c^{3/2}}{(b-a)(c-a)} \,
    \bigsbrk{ \delta_{ad} \delta_{be} \delta_{cf}
            - \delta_{ad} \delta_{bf} \delta_{ce}} \nl
  + \frac{\g^2 a^{3/2}b^{2}c^{3/2}\bigsbrk{(b-a)c-a(b+c)}}{(b-a)(c-a)(b+c)} \,
    \bigsbrk{ \delta_{ae} \delta_{bd} \delta_{cf}
            - \delta_{af} \delta_{bd} \delta_{ce}} \nl
  + \frac{\g^2a^{3/2}b^{3/2}c^{3}}{(a+c)(b+c)} \,
    \bigsbrk{ \delta_{ae} \delta_{bf} \delta_{cd}
            - \delta_{af} \delta_{be} \delta_{cd}} \; ,
\\[3mm]
\Amp^{\mathrm{1-loop}}(\pin) \eq
  + \frac{i\g^3a^4b^{3/2}c^{3/2}[c(b-c)+a(b+c)]}{(b-a)(c^2-a^2)} \,
    \bigsbrk{ \delta_{ad} \delta_{be} \delta_{cf}
            - \delta_{ad} \delta_{bf} \delta_{ce}} \nl
  + \frac{i\g^3a^{3/2}b^{2}c^{3/2}}{(b-a)(c^2-b^2)(c^2-a^2)}
    \biggsbrk{2a^2c^3(a+c)+2abc^2(a-c)(a+2c)\nl
    \quad -b^3(a-c)(a+c)^2-b^2c(3a^3+a^2c+ac^2-c^3)} \,
    \bigsbrk{ \delta_{ae} \delta_{bd} \delta_{cf}
            - \delta_{af} \delta_{bd} \delta_{ce}} \nl
  + \frac{i\g^3a^{3/2}b^{3/2}c^4[a(b-c)+b(b+c)]}{(c^2-b^2)(a+c)} \,
    \bigsbrk{ \delta_{ae} \delta_{bf} \delta_{cd}
            - \delta_{af} \delta_{be} \delta_{cd}}
\ee
This difference comes about as follows. Firstly, the exchange of the in-coming and out-going particles reshuffles the momenta $\{a,b,c\}$ in the matrix elements. Secondly, the matrix elements are simplified making use of the initial state condition $c>b>a>0$. This makes different matrix elements incomparable. The partial amplitudes $\Amp^{\mathrm{dog}}$ and $\Amp^{\mathrm{sun}}$ are printed in \appref{app:scatt-amp}.

%%%%%%%%%%%%%%%%%%%%%%%%%%%%%%%%%%%%%%%%%%%%%%%%%%%%%%%%%%%%%%%%%%%%%%%%%%%
%%%%%%%%%%%%%%%%%%%%%%%%%%%%%%%%%%%%%%%%%%%%%%%%%%%%%%%%%%%%%%%%%%%%%%%%%%%
\section{Factorization of the three-particle S-matrix}
\label{sec:factorization}

We now show that the three-particle S-matrix elements computed in the previous section are indeed products of corresponding elements of the two-particle S-matrix recorded in \secref{sec:MS-model}. Due to the high degree of symmetry this is sufficient to prove the factorization of the full three-particle S-matrix to one-loop order (i.e. order $\g^3$).

Specifically, we verify the factorization by calculating the triple product of two-particle S-matrices according to \eqref{eqn:Smat-factorization} and showing that this product agrees with the computed three-particle amplitudes. It will be instructive to not only compare the total amplitudes but certain contributions individually. To this end we split the tree-level and one-loop S-matrix elements as follows
\[
\begin{split} \label{eqn:Smat-split}
\Smatrix^{(0)} & = \Smatrix^{11\g^2} + \Smatrix^{1\g\g} \; , \\
\Smatrix^{(1)} & = \Smatrix^{11\g^3} + \Smatrix^{1\g\g^2} + \Smatrix^{\g\g\g} \; ,
\end{split}
\]
where the superscripts indicate the perturbative order of the three factors in \eqref{eqn:Smat-factorization}. For instance, $\Smatrix^{1\g\g^2}$ refers to all terms in the triple product that originate from taking the zeroth order in $\g$ from one of the three two-particle S-matrices, the first order from one of the remaining S-matrices and the second order from the final S-matrix.

The first terms in the equations \eqref{eqn:Smat-split} describe processes where one of the particles does not take part in the interaction. These are precisely the terms that correspond to \emph{disconnected} Feynman diagrams. Since we omitted them in the computation in \secref{sec:3p-scatt}, we have to discard these terms here, too. We were allowed to disregard these contributions because their factorization is trivial; see the discussion at the end of \secref{sec:symmetries}.

The structure of the one-loop diagrams, \figref{fig:one-loop} on page \pageref{fig:one-loop}, suggests the correspondence
\[ \label{eqn:one-loop-structure}
 \Smatrix^{1\g\g^2} \stackrel{?}{\longleftrightarrow} \Amp^{\mathrm{dog}}
 \comma
 \Smatrix^{\g\g\g} \stackrel{?}{\longleftrightarrow} \Amp^{\mathrm{sun}} \; .
\]
While \eqref{eqn:one-loop-structure} is valid for some special amplitudes, it turns out \emph{not} to be true in general. In fact, there is an important conceptual difference between the factorized S-matrix and the Feynman diagrams. On the one hand, the factorized S-matrix is a series of two-particle scatterings of physical particles, i.e.\ the particles propagating between the interaction points are on-shell. On the other hand, the particles propagating insides the Feynman diagrams are virtual. Therefore, there is no reason why the correspondence \eqref{eqn:one-loop-structure} should hold. All that really matters for factorization is that the sums of the terms in \eqref{eqn:one-loop-structure} match.

Let us now look at the processes computed in the last section in detail.

%%%%%%%%%%%%%%%%%%%%%%%%%%%%%%%%%%%%%%%%%%%%%%%%%%%%%%%%%%%%%%%%%%%%%%%%%%%
\subsection{Highest weight processes}

%%%%%%%%%%%%%%%%%%%%%%%%%%%%%%%%%%%%%%%%%%%%%%%%%%%%%%%%%%%%%%%%%%%%%%%%%%%
\subsubsection*{$Y_{1\dot{1}}(a) Y_{1\dot{1}}(b) Y_{1\dot{1}}(c)
     \rightarrow Y_{1\dot{1}}(a) Y_{1\dot{1}}(b) Y_{1\dot{1}}(c)$}

Using the near-flat-space S-matrix from \secref{sec:MS-model} in the factorization equation \eqref{eqn:Smat-factorization}, we find for the three-particle S-matrix element governing this process:
\[ \label{eqn:11full}
\Smatrix_{1\dot{1}}^{\mathrm{full}}(a,b,c) =
\bigeval{S_0(A+B)^2}_{(a,b)} \,
\bigeval{S_0(A+B)^2}_{(a,c)} \,
\bigeval{S_0(A+B)^2}_{(b,c)} \; ,
\]
where the relevant coefficients from \eqref{eqn:Smat-coeffs} are
\[ \label{eqn:A+B}
\bigeval{(A+B)}_{(a,b)} = 1 + i\g a b \frac{b+a}{b-a} \; .
\]
This sum corresponds to those terms in the two-particle S-matrix which symmetrize two bosonic indices. Expanding the matrix element \eqref{eqn:11full} in $\g$, one finds the prediction for the connected tree-level amplitude
\be
\Smatrix_{1\dot{1}}^{1\g\g} \eq -4\,\g^2\,a\,b\,c
\lrsbrk{
a\,\frac{(a+b)(a+c)}{(a-b)(a-c)} +
b\,\frac{(a+b)(b+c)}{(a-b)(b-c)} +
c\,\frac{(a+c)(b+c)}{(a-c)(b-c)}
}
\ee
and the two pieces of the one-loop amplitude
\be
\Smatrix_{1\dot{1}}^{1\g\g^2} \eq
-\!\!\sum_{\sigma(a,b,c)} \frac{4i\g^3 a^3 b^2 c}{a^2-b^2}
\lrabs{\frac{a+c}{a-c}}
\biggsbrk{
\frac{(a+b)^3}{a-b} \nl
\hspace{20mm}-\frac{4i}{\pi} \frac{ab}{\lrabs{a^2-b^2}}
\lrbrk{a^2-b^2+\lrbrk{a^2+b^2}\ln\frac{b}{a}}
} \; , \\[3mm]
\Smatrix_{1\dot{1}}^{\g\g\g} \eq
8i\,\g^3\,a^2b^2c^2 \frac{(a+b)(a+c)(b+c)}{(a-b)(a-c)(b-c)} \; .
\ee
These results match those from the Feynman diagrams, in particular we see that in this case the correspondence \eqref{eqn:one-loop-structure} does hold.

%%%%%%%%%%%%%%%%%%%%%%%%%%%%%%%%%%%%%%%%%%%%%%%%%%%%%%%%%%%%%%%%%%%%%%%%%%%
\subsubsection*{$Z_{3\dot{3}}(a) Z_{3\dot{3}}(b) Z_{3\dot{3}}(c)
     \rightarrow Z_{3\dot{3}}(a) Z_{3\dot{3}}(b) Z_{3\dot{3}}(c)$}

The full matrix element is given by
\[
\Smatrix_{3\dot{3}}^{\mathrm{full}}(a,b,c) =
\bigeval{S_0(D+E)^2}_{(a,b)} \,
\bigeval{S_0(D+E)^2}_{(a,c)} \,
\bigeval{S_0(D+E)^2}_{(b,c)} \; ,
\]
where the sum of coefficients, corresponding to the symmetrizer of two fermionic indices, is given by
\[
\bigeval{(D+E)}_{(a,b)} = 1 - i\g a b \frac{b+a}{b-a} \; .
\]
Since the phase factor $S_0$ is even in $\g$ to the considered order, one immediately sees that the S-matrix elements for this process can be obtained from the ones for the scattering of $Y_{1\dot{1}}$'s by sending $\g\to-\g$. Hence, we get
\[
  \Smatrix_{3\dot{3}}^{1\g\g} = \Smatrix_{1\dot{1}}^{1\g\g}
  \comma
  \Smatrix_{3\dot{3}}^{1\g\g^2} = -\Smatrix_{1\dot{1}}^{1\g\g^2}
  \comma
  \Smatrix_{3\dot{3}}^{\g\g\g} = -\Smatrix_{1\dot{1}}^{\g\g\g} \; .
\]
This is the same relation between the Feynman amplitudes involving $Y$- and $Z$-bosons in \secref{sec:essential-processes}. Thus also this prediction agrees with the diagrammatic calculation.

%%%%%%%%%%%%%%%%%%%%%%%%%%%%%%%%%%%%%%%%%%%%%%%%%%%%%%%%%%%%%%%%%%%%%%%%%%%
\subsubsection*{$\Psi_{1\dot{3}}(a) \Psi_{1\dot{3}}(b) \Psi_{1\dot{3}}(c)
     \rightarrow \Psi_{1\dot{3}}(a) \Psi_{1\dot{3}}(b) \Psi_{1\dot{3}}(c)$}

The full amplitude for this process is
\[
\Smatrix_{1\dot{3}}^{\mathrm{full}}(a,b,c) =
\bigeval{S_0(A+B)(D+E)}_{(a,b)} \,
\bigeval{S_0(A+B)(D+E)}_{(a,c)} \,
\bigeval{S_0(A+B)(D+E)}_{(b,c)} \; .
\]
Writing out the momentum dependence we get
\[
\bigeval{(A+B)(D+E)}_{(a,b)} =
1 + \g^2 a^2 b^2 \lrbrk{\frac{a+b}{a-b}}^2.
\]
This factor exactly cancels the denominator of $S_0(a,b)$. Hence the full amplitude for this process is given by
\[
\begin{split}
\Smatrix_{1\dot{3}}^{\mathrm{full}} = 1 - \frac{8i\g^2}{\pi}
\biggsbrk{
  &\ \frac{a^3b^3}{(a^2-b^2)^2}
  \biggbrk{a^2-b^2+(a^2+b^2)\ln\frac{b}{a}} \\
+ &\ \frac{a^3c^3}{(a^2-c^2)^2}
  \biggbrk{a^2-c^2+(a^2+c^2)\ln\frac{c}{a}} \\
+ &\ \frac{b^3c^3}{(b^2-c^2)^2}
  \biggbrk{b^2-c^2+(b^2+c^2)\ln\frac{c}{b}} }
+ \order(\g^4) \; .
\end{split}
\]
From this amplitude, we see that the lowest order connected part is of order $\g^4$, so that
\[
  \Smatrix_{1\dot{3}}^{1\g\g} = 0
  \comma
  \Smatrix_{1\dot{3}}^{1\g\g^2} = 0
  \comma
  \Smatrix_{1\dot{3}}^{\g\g\g} = 0 \; .
\]
Also the Feynman diagram calculation resulted in vanishing tree-level and one-loop amplitudes.

%%%%%%%%%%%%%%%%%%%%%%%%%%%%%%%%%%%%%%%%%%%%%%%%%%%%%%%%%%%%%%%%%%%%%%%%%%%
\subsubsection*{$\Bsi_{3\dot{1}}(a) \Bsi_{3\dot{1}}(b) \Bsi_{3\dot{1}}(c)
     \rightarrow \Bsi_{3\dot{1}}(a) \Bsi_{3\dot{1}}(b) \Bsi_{3\dot{1}}(c)$}

The S-matrix element describing this process can be obtained from the previous one, by exchanging undotted and dotted indices. Using \eqref{eqn:Smat-non-symmetry}, we get
\be
  \Smatrix_{3\dot{1}}^{\mathrm{full}} = \Smatrix_{1\dot{3}}^{\mathrm{full}}
  \comma
  \Smatrix_{3\dot{1}}^{1\g\g} = 0
  \comma
  \Smatrix_{3\dot{1}}^{1\g\g^2} = 0
  \comma
  \Smatrix_{3\dot{1}}^{\g\g\g} = 0 \; ,
\ee
a result that once again agrees with the perturbative calculation. This also shows that factorization of the three-particle S-matrix predicts $C_2=C_3$ to hold exactly in \eqref{eqn:S-matrix-proj}.

%%%%%%%%%%%%%%%%%%%%%%%%%%%%%%%%%%%%%%%%%%%%%%%%%%%%%%%%%%%%%%%%%%%%%%%%%%%
\subsection{Mixed processes}
\label{sec:bonus-prediction}

In \secref{sec:bonus-processes}, we have considered the amplitudes for some processes where fields from all four $\algSU(2)$ sectors are involved. Here, we show the factorization of all S-matrix elements that can be read off from those amplitudes.

%%%%%%%%%%%%%%%%%%%%%%%%%%%%%%%%%%%%%%%%%%%%%%%%%%%%%%%%%%%%%%%%%%%%%%%%%%%
\subsubsection*{$Y_1(a) \psi_1(b) \psi_6(c) \rightarrow Z_7(a) Y_2(b) Y_3(c) + \ldots $}

To show the factorization of the corresponding amplitudes, we need to compute the S-matrix elements for the scattering of $Y_1(a) \psi_1(b) \psi_6(c)$ into all different orderings of $Z_7\, Y_2\, Y_3$. Restricting the S-matrix to such out-states, we find
\begin{align}
&\Smatrix^{1\g\g}\ket{Y_1(a)\,\psi_1(b)\,\psi_6(c)} = \\[2mm]
  &-\frac{\g^2 a^{3}b^{3/2}c^{3/2}}{(b-a)(c-a)} \, \Bigbrk{
       \ket{Z_7(a)\,Y_2(b)\,Y_3(c)}
     - \ket{Z_7(a)\,Y_3(b)\,Y_2(c)} \nn \\
  & \hspace{29mm}
     + \ket{Y_2(a)\,Z_7(b)\,Y_3(c)}
     - \ket{Y_3(a)\,Z_7(b)\,Y_2(c)} } \nn \\
  &+\frac{\g^2 a^{3}b^{3/2}c^{3/2}}{(a+b)(a+c)} \, \Bigbrk{
       \ket{Y_2(a)\,Y_3(b)\,Z_7(c)}
     - \ket{Y_3(a)\,Y_2(b)\,Z_7(c)} } \; , \nn \\[3mm]
&\Smatrix^{1\g\g^2}\ket{Y_1(a)\,\psi_1(b)\,\psi_6(c)} = \label{eqn:mixed-012} \\[2mm]
  &+\frac{i\g^3a^4b^{3/2}c^{3/2}\bigsbrk{(b-c)c+a(b+c)}}{(b-a)(c^2-a^2)} \, \Bigbrk{
      \ket{Z_7(a)\,Y_2(b)\,Y_3(c)}
    - \ket{Z_7(a)\,Y_3(b)\,Y_2(c)}} \nn \\
  &-\frac{i\g^3a^4b^{3/2}c^{3/2}\bigsbrk{a(b-c)+c(b+c)}}{(b-a)(c^2-a^2)} \, \Bigbrk{
      \ket{Y_2(a)\,Z_7(b)\,Y_3(c)}
    - \ket{Y_3(a)\,Z_7(b)\,Y_2(c)}} \nn \\
  &+\frac{i\g^3a^4b^{3/2}c^{3/2}\bigsbrk{a(b-c)+b(b+c)}}{(b^2-a^2)(a+c)} \, \Bigbrk{
      \ket{Y_2(a)\,Y_3(b)\,Z_7(c)}
    - \ket{Y_3(a)\,Y_2(b)\,Z_7(c)}} \; , \nn \\[3mm]
&\Smatrix^{\g\g\g}\ket{Y_1(a)\,\psi_1(b)\,\psi_6(c)} = 0 \label{eqn:mixed-111} \; .
\end{align}
We note that all coefficients \eqref{eqn:Smat-coeffs} of the two-particle S-matrix except $D$ and $E$ enter these expressions. The agreement with the amplitudes \eqref{eqn:mixed-forward} is perfect, as long as we consider the total one-loop result. However, when we try to compare \eqref{eqn:mixed-012} and \eqref{eqn:mixed-111} to the partial amplitudes \eqref{eqn:mixed-dog} and \eqref{eqn:mixed-sun} separately, we see that the hypothetical identification \eqref{eqn:one-loop-structure} fails for these more complicated processes.

%%%%%%%%%%%%%%%%%%%%%%%%%%%%%%%%%%%%%%%%%%%%%%%%%%%%%%%%%%%%%%%%%%%%%%%%%%%
\subsubsection*{$Z_7(a) Y_2(b) Y_3(c) \rightarrow Y_1(a) \psi_1(b) \psi_6(c) + \ldots $}

We avoid to print the prediction of the corresponding S-matrix elements, but simply report the agreement with \eqref{eqn:mixed-reversed}. In this case all of the coefficients from \eqref{eqn:Smat-coeffs} are probed. Here the contribution from $\Smatrix^{\g\g\g}$ is non-zero, nevertheless the correspondence \eqref{eqn:one-loop-structure} does not hold here either.

%%%%%%%%%%%%%%%%%%%%%%%%%%%%%%%%%%%%%%%%%%%%%%%%%%%%%%%%%%%%%%%%%%%%%%%%%%%
%%%%%%%%%%%%%%%%%%%%%%%%%%%%%%%%%%%%%%%%%%%%%%%%%%%%%%%%%%%%%%%%%%%%%%%%%%%
\section{Conclusions and outlook}

We have studied three-particle scattering amplitudes in the decompactified world-sheet theory describing superstrings in $\AdS_5\times\Sphere^5$ in the near-flat limit. The three-particle S-matrix governing these processes was seen to be restricted by the $\algPSU(2|2)^2\ltimes\Reals^3$ symmetry of the model up to 4 scalar functions, recall \eqref{eqn:S-matrix-proj}. We computed a set of one-loop amplitudes (\secref{sec:3p-scatt}) which determine these functions up to order $\g^3 \sim \frac{1}{\lambda^{3/2}}$. Subsequently, we showed that the corresponding three-particle S-matrix elements can be expressed in terms of the model's two-particle S-matrix according to the Zamolodchikov algebra (\secref{sec:factorization}). On the basis of the $\algPSU(2|2)^2\ltimes\Reals^3$ symmetry, this proves the factorization of the entire three-particle world-sheet S-matrix of strings in near-flat $\AdS_5\times\Sphere^5$ at the first quantum level.

In addition to the highest weight processes which fix the freedom in the S-matrix in the most direct way, we have computed some more complicated, flavor changing amplitudes in \secref{sec:bonus-processes} and \ref{sec:bonus-prediction}. This serves as an indirect check of the supersymmetries of the near-flat-space model. It also helps to get insight into the mechanism of factorization from a diagrammatic point of view. In the highly symmetric cases it is possible to regard the factorization equation \figref{fig:factorization} (page \pageref{fig:factorization}) directly as Feynman diagrams with grouped vertices. In general, however, the correspondence of seemingly equivalent structures in the perturbation expansion of the factorization equation and the series of Feynman graphs as in \eqref{eqn:one-loop-structure} does not hold. One may consider \figref{fig:factorization} at best as effective diagrams after many non-trivial cancelations between different Feynman graphs have taken place. These cancelations are a consequence of the higher conservation laws, but occur completely unheralded in the diagrammatic computation. It would be very interesting to extend these considerations to two-loop order, where there are Feynman diagrams which cannot even be drawn as a sequence of three two-particle interactions, see e.g. \figref{fig:generic2loop} on page \pageref{fig:generic2loop}. Moreover, for two-loop diagrams the cutting rule of \cite{Kallen:1965}, which relates one-loop graphs to dressed tree-level graphs, does not apply straightforwardly. Thus, one should investigate how the poles in the amplitudes can be extracted at higher loop order.

Our computation provides an explicit and direct check of the quantum integrability of the near-flat-space model describing a single string in the Maldacena-Swanson limit of $\AdS_5\times\Sphere^5$ \cite{Maldacena:2006rv}. A complete proof of integrability would, of course, require showing the factorization of all multi-particle S-matrices to all orders (or better non-perturbatively). This is very hard to do by direct means but certainly not impossible, see e.g. \cite{Thacker:1974kv, Thacker:1976vp}. However, already the existence of the first higher conserved charge, which follows from the factorization of three-particle scattering, is under certain conditions sufficient for quantum integrability \cite{Parke:1980ki, Grabowski:1994rb}.

We have used general arguments from the representation theory of $\algPSU(2|2)^2\ltimes\Reals^3$ \cite{Beisert:2006qh} to relate the various elements of the three-particle S-matrix and show the factorization from a minimal number of amplitudes. To render our proof mathematically rigorous, one should, however, find the action of the symmetry generators explicitly. Their form is not known for the near-flat-space model to date and taking the corresponding limit of the generators for the full AdS string \cite{Arutyunov:2006ak} is not straightforward due to non-local field redefinitions.

It would also be interesting to go beyond these symmetries and investigate the near-flat-space model with respect to Yangian symmetry \cite{Drinfeld:1985rx}. This symmetry would restrict the S-matrix even stronger and leave only one functional degree of freedom for a given particle number \cite{Beisert:2007ds}. In fact, one could work out the action of the Yangian generators explicitly and verify that the four functions which we have used to specify the three-particle S-matrix are indeed related.

%%%%%%%%%%%%%%%%%%%%%%%%%%%%%%%%%%%%%%%%%%%%%%%%%%%%%%%%%%%%%%%%%%%%%%%%%%%
\bigskip
\subsection*{Acknowledgments}
\bigskip
We thank J.~Minahan for suggesting the problem and we are grateful to him and K.~Zarembo, N.~Beisert, T.~McLoughlin and M.~Staudacher for many interesting and helpful discussions. Moreover, we are thankful to J.~Minahan and K.~Zarembo for their valuable comments on the manuscript. The work of T.K. was supported by the G\"oran Gustafsson Foundation. V.G.M.P. would like to thank VR for partial financial support.
% under the grant .....

\appendix

%%%%%%%%%%%%%%%%%%%%%%%%%%%%%%%%%%%%%%%%%%%%%%%%%%%%%%%%%%%%%%%%%%%%%%%%%%%
\section{Notation} \label{app:notation}

\begin{figure}
\begin{center}
\includegraphics[scale=0.6]{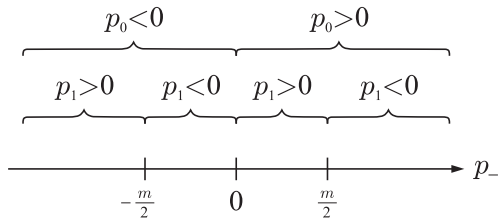}
\end{center}
\caption{\textbf{Light-cone momentum.} Depending on the value of $p_-$, the corresponding excitation is a right-moving anti-particle, a left-moving anti-particle, a right-moving particle or a left-moving particle.}
\label{fig:lightconemomentum}
\end{figure}

On the world-sheet, we use the metric $\eta_{\mu\nu} = (+,-)$ and introduce light-cone coordinates and momenta as
\[ \label{eqn:lc-momenta}
\sigma^\pm = \sigma^0 \pm \sigma^1
\comma
p_\pm = \half (p_0 \pm p_1) \; .
\]
The parametrization of scattering amplitudes in terms of the light-cone momentum $p_-$ is very convenient as it covers all possibilities, particle/anti-particle and left-/right-mover, with one single real variable. The different cases are sketched in \figref{fig:lightconemomentum}.
Since we are dealing with many different momenta it is important to simplify the notation. We will denote the minus component of a momentum simply by the momentum label, e.g. $a \equiv p_{a-}$. When we refer to both components, we use $\vec{a} \equiv (p_{a-},p_{a+})$. The amplitudes are functions of six momenta $a$ to $f$. We introduce shorthands for two special sequences that occur frequently $\pin = (a,b,c,d,e,f)$ and $\pou = (a,b,c,-d,-e,-f)$.

The way in which the near-flat-space model treats left- and right-movers makes it convenient to perform the quantization in world-sheet light-cone coordinates with $\sigma^+$ as time. Thus the target-space fields have the mode expansions
\begin{align}
Y_{i'}(\vecsigma) & = \int\frac{dp_-}{2\pi} \frac{1}{\sqrt{2p_-}} \:
                 \Bigsbrk{ a_{i'}(p_-)      \, e^{-i\vec{p}\cdot\vecsigma}
                         + a^\dag_{i'}(p_-) \, e^{+i\vec{p}\cdot\vecsigma} } \; , \\
Z_i(\vecsigma)    & = \int\frac{dp_-}{2\pi} \frac{1}{\sqrt{2p_-}} \:
                    \Bigsbrk{ a_i(p_-)      \, e^{-i\vec{p}\cdot\vecsigma}
                            + a^\dag_i(p_-) \, e^{+i\vec{p}\cdot\vecsigma} } \; , \\
\psi(\vecsigma)   & = \int\frac{dp_-}{2\pi} \frac{1}{\sqrt{2}} \:
                    \Bigsbrk{ b(p_-)        \, e^{-i\vec{p}\cdot\vecsigma}
                            + b^\dag(p_-)   \, e^{+i\vec{p}\cdot\vecsigma} } \; .
\end{align}
\noindent
The free bosonic and fermionic propagators are
\[
\frac{i}{\vec{p}^2 - m^2} \comma \frac{ip_-}{\vec{p}^2 - m^2} \; ,
\]
and the free dispersion relation is $2 p_+ =\tfrac {m^2}{2 p_-} $.

We use the following representation for the $16\times 16$ Dirac-matrices
\newcommand{\ourtimes}{}
\begin{align} \label{eqn:Diracmat}
\Gamma_{1} & = \half \brk{- \sigma_{2} \ourtimes \sigma_{0} \ourtimes \sigma_{2} \ourtimes \sigma_{0}
                          - \sigma_{2} \ourtimes \sigma_{0} \ourtimes \sigma_{2} \ourtimes \sigma_{3}
                          - \sigma_{2} \ourtimes \sigma_{3} \ourtimes \sigma_{2} \ourtimes \sigma_{0}
                          + \sigma_{2} \ourtimes \sigma_{3} \ourtimes \sigma_{2} \ourtimes \sigma_{3} } \; , \\
\Gamma_{2} & = \half \brk{+ \sigma_{2} \ourtimes \sigma_{0} \ourtimes \sigma_{1} \ourtimes \sigma_{2}
                          - \sigma_{2} \ourtimes \sigma_{0} \ourtimes \sigma_{2} \ourtimes \sigma_{1}
                          + \sigma_{2} \ourtimes \sigma_{3} \ourtimes \sigma_{1} \ourtimes \sigma_{2}
                          + \sigma_{2} \ourtimes \sigma_{3} \ourtimes \sigma_{2} \ourtimes \sigma_{1} } \; , \nn \\
\Gamma_{3} & = \hspace{7mm} \sigma_{1} \ourtimes \sigma_{0} \ourtimes \sigma_{0} \ourtimes \sigma_{0}   \; , \nn \\
\Gamma_{4} & = \half \brk{+ \sigma_{2} \ourtimes \sigma_{0} \ourtimes \sigma_{0} \ourtimes \sigma_{2}
                          - \sigma_{2} \ourtimes \sigma_{0} \ourtimes \sigma_{3} \ourtimes \sigma_{2}
                          - \sigma_{2} \ourtimes \sigma_{3} \ourtimes \sigma_{0} \ourtimes \sigma_{2}
                          - \sigma_{2} \ourtimes \sigma_{3} \ourtimes \sigma_{3} \ourtimes \sigma_{2} } \; , \nn \\
\Gamma_{5} & = \half \brk{- \sigma_{2} \ourtimes \sigma_{1} \ourtimes \sigma_{2} \ourtimes \sigma_{0}
                          + \sigma_{2} \ourtimes \sigma_{1} \ourtimes \sigma_{2} \ourtimes \sigma_{3}
                          - \sigma_{2} \ourtimes \sigma_{2} \ourtimes \sigma_{1} \ourtimes \sigma_{0}
                          - \sigma_{2} \ourtimes \sigma_{2} \ourtimes \sigma_{1} \ourtimes \sigma_{3} } \; , \nn \\
\Gamma_{6} & = \half \brk{+ \sigma_{2} \ourtimes \sigma_{1} \ourtimes \sigma_{1} \ourtimes \sigma_{2}
                          + \sigma_{2} \ourtimes \sigma_{1} \ourtimes \sigma_{2} \ourtimes \sigma_{1}
                          - \sigma_{2} \ourtimes \sigma_{2} \ourtimes \sigma_{1} \ourtimes \sigma_{1}
                          - \sigma_{2} \ourtimes \sigma_{2} \ourtimes \sigma_{2} \ourtimes \sigma_{2} } \; , \nn \\
\Gamma_{7} & = \half \brk{+ \sigma_{2} \ourtimes \sigma_{2} \ourtimes \sigma_{0} \ourtimes \sigma_{0}
                          - \sigma_{2} \ourtimes \sigma_{2} \ourtimes \sigma_{0} \ourtimes \sigma_{3}
                          + \sigma_{2} \ourtimes \sigma_{2} \ourtimes \sigma_{3} \ourtimes \sigma_{0}
                          + \sigma_{2} \ourtimes \sigma_{2} \ourtimes \sigma_{3} \ourtimes \sigma_{3} } \; , \nn \\
\Gamma_{8} & = \half \brk{- \sigma_{2} \ourtimes \sigma_{1} \ourtimes \sigma_{0} \ourtimes \sigma_{2}
                          - \sigma_{2} \ourtimes \sigma_{1} \ourtimes \sigma_{3} \ourtimes \sigma_{2}
                          - \sigma_{2} \ourtimes \sigma_{2} \ourtimes \sigma_{0} \ourtimes \sigma_{1}
                          + \sigma_{2} \ourtimes \sigma_{2} \ourtimes \sigma_{3} \ourtimes \sigma_{1} } \; , \nn
\end{align}
where the sequence of $\sigma$'s denotes the outer product of the $2\times2$ identity $\sigma_0$ and the standard Pauli matrices $\sigma_{1,2,3}$. We also define
\[
\Pi = \Gamma^1 \Gamma^2 \Gamma^3 \Gamma^4
\comma
\Gamma^9 = \Gamma^1 \Gamma^2 \cdots \Gamma^8
\comma
P_{R,L} = \half(\unit\pm\Gamma^9) \; .
\]

%%%%%%%%%%%%%%%%%%%%%%%%%%%%%%%%%%%%%%%%%%%%%%%%%%%%%%%%%%%%%%%%%%%%%%%%%%%
\section{Embedding of $\algSU(2)^4$ into $\algSO(8)$}
\label{app:embedding}

We define the embedding of the four $\algSU(2)$ subalgebras into $\algSO(8)$ through the following identification of the $\algSU(2)$ generators $J$,$\dot{J}$,$S$,$\dot{S}$ with the rotation $\algSO(8)$ generators $R_{IJ}$:
\begin{align}
 J_1       & = \half \brk{+R_{12} - R_{34}} \; , &
 \dot{J}_1 & = \half \brk{+R_{12} + R_{34}} \; , \nln
 J_2       & = \half \brk{+R_{13} + R_{24}} \; , &
 \dot{J}_2 & = \half \brk{-R_{13} + R_{24}} \; , \nln
 J_3       & = \half \brk{+R_{14} - R_{23}} \; , &
 \dot{J}_3 & = \half \brk{+R_{14} + R_{23}} \; , \label{eqn:so8-su2} \\[4mm]
 S_1       & = \half \brk{+R_{56} - R_{78}} \; , &
 \dot{S}_1 & = \half \brk{+R_{56} + R_{78}} \; , \nln
 S_2       & = \half \brk{+R_{57} + R_{68}} \; , &
 \dot{S}_2 & = \half \brk{-R_{57} + R_{68}} \; , \nln
 S_3       & = \half \brk{+R_{58} - R_{67}} \; , &
 \dot{S}_3 & = \half \brk{+R_{58} + R_{67}} \; . \nn
\end{align}
Raising and lowering generators are defined in the standard way, $J_\pm = J_1 \pm i J_2$, and similar for the other $\algSU(2)$'s. The fields $Y_{\lAA\rAA}$, $Z_{\laa\raa}$, $\Psi_{\lAA\raa}$, $\Bsi_{\laa\rAA}$ are defined as the components of the vectors $Y_{i'}$, $Z_i$ and the spinors $\psi_\pm$ with certain $J_3$,$\dot{J}_3$,$S_3$,$\dot{S}_3$ charges, cf. \tabref{tab:fields-indices}.
\begin{table}
\begin{center}
\begin{tabular}{|c|c|c|c|c|} \cline{2-5}
\multicolumn{1}{c|}{} & \multicolumn{2}{c|}{$\Sphere^5$} &
\multicolumn{2}{c|}{$\AdS_5$ \vphantom{\rule[-2mm]{0mm}{7mm}}} \\
\hline
                  & $\grSU(2)$  & $\grSU(2)$  & $\grSU(2)$  & $\grSU(2)$  \vphantom{\rule[-2mm]{0mm}{7mm}} \\ \hline
``Spin''          & $J$         & $\dot{J}$   & $\dot{S}$   & $S$         \vphantom{\rule{0mm}{5mm}} \\
Index             & $\lAA = 1,2$
                  & $\rAA = \dot{1},\dot{2}$
                  & $\raa = \dot{3},\dot{4}$
                  & $\laa = 3,4$                                          \\ \hline
$Y_{\lAA\rAA}$    & $\rep{2}$   & $\rep{2}$   & $\rep{1}$   & $\rep{1}$   \vphantom{\rule{0mm}{5mm}} \\
$Z_{\laa\raa}$    & $\rep{1}$   & $\rep{1}$   & $\rep{2}$   & $\rep{2}$   \\
$\Psi_{\lAA\raa}$ & $\rep{2}$   & $\rep{1}$   & $\rep{2}$   & $\rep{1}$   \\
$\Bsi_{\laa\rAA}$ & $\rep{1}$   & $\rep{2}$   & $\rep{1}$   &
$\rep{2}$   \\ \hline
\end{tabular}
\end{center}
\caption{\textbf{\mathversion{bold}$\grSU(2)^4$ quantum numbers.}
% Representations of $\grSU(2)^4$ will be denoted by $(\rep{2J+1 , 2\dot{J}+1 , 2\dot{S}+1, 2S+1})$.
}
\label{tab:fields-indices}
\end{table}
To construct the bosons, one uses the 8-dimensional vector representation
\[
R_{IJ} \rightarrow M_{IJ}
\comma
(M_{IJ})_{KL} = i \lrbrk{ \delta_{IK} \delta_{JL} - \delta_{JK} \delta_{IL} } \; .
\]
We find%
\footnote{This is the same as the identification $Y_{\lAA\rAA} = \rho_{i'} Y_{i'}$, $Z_{\laa\raa} = \rho_i Z_i$ with $\rho_i = (\unit,i\sigma_i)$ used in \cite{Klose:2007rz} and \cite{Klose:2006zd}. In fact \eqref{eqn:so8-su2} has been determined such that this identification holds.}
\begin{align}
  Y_1 & = + \half (Y_{1\dot{1}} + Y_{2\dot{2}}) \; , &
  Z_5 & = + \half (Z_{3\dot{3}} + Z_{4\dot{4}}) \; , \nln
  Y_2 & = -\ihalf (Y_{1\dot{2}} + Y_{2\dot{1}}) \; , &
  Z_6 & = -\ihalf (Z_{3\dot{4}} + Z_{4\dot{3}}) \; , \label{eqn:su2-boso} \\
  Y_3 & = + \half (Y_{1\dot{2}} - Y_{2\dot{1}}) \; , &
  Z_7 & = + \half (Z_{3\dot{4}} - Z_{4\dot{3}}) \; , \nln
  Y_4 & = -\ihalf (Y_{1\dot{1}} - Y_{2\dot{2}}) \; , &
  Z_8 & = -\ihalf (Z_{3\dot{3}} - Z_{4\dot{4}}) \; . \nn
\end{align}
The free action for the bosons in the two different notations read
\be \label{eqn:free-boson-action}
\Lagr_0^{\mathrm{bosons}}
  \eq 2 \, \partial_+ Y_{i'} \, \partial_- Y_{i'} - \tfrac{m^2}{2} \, Y_{i'} \, Y_{i'}
    + 2 \, \partial_+ Z_{i } \, \partial_- Z_{i } - \tfrac{m^2}{2} \, Z_{i } \, Z_{i } \\
          \eq + \tfrac{1}{4} \tim{Y}^*_{\lAA\rAA} \tim{Y}^{\lAA\rAA}
              - \tfrac{1}{4} \spa{Y}^*_{\lAA\rAA} \spa{Y}^{\lAA\rAA}
              - \tfrac{m^2}{4}   {Y}^*_{\lAA\rAA}     {Y}^{\lAA\rAA} \nl
              + \tfrac{1}{4} \tim{Z}^*_{\laa\raa} \tim{Z}^{\laa\raa}
              - \tfrac{1}{4} \spa{Z}^*_{\laa\raa} \spa{Z}^{\laa\raa}
              - \tfrac{m^2}{4}   {Z}^*_{\laa\raa}     {Z}^{\laa\raa} \nn \; .
\ee

To construct the fermions, one uses the spinor representation
\[
  R_{IJ} \rightarrow \ihalf \Gamma_{IJ}
  \comma
  \Gamma_{IJ} = \half \comm{\Gamma_I}{\Gamma_J} \; .
\]
Starting with a complex, left-chiral spinor $\Theta$, we find the components
\begin{align}
\Theta_1 & = + \half (\Psi_{1\dot{3}} + \Psi_{2\dot{4}}) \; , &
\Theta_5 & = + \half (\Bsi_{3\dot{1}} + \Bsi_{4\dot{2}}) \; , \nln
\Theta_2 & = -\ihalf (\Psi_{1\dot{4}} + \Psi_{2\dot{3}}) \; , &
\Theta_6 & = -\ihalf (\Bsi_{3\dot{2}} + \Bsi_{4\dot{1}}) \; , \label{eqn:su2-ferm} \\
\Theta_3 & = + \half (\Psi_{1\dot{4}} - \Psi_{2\dot{3}}) \; , &
\Theta_7 & = + \half (\Bsi_{3\dot{2}} - \Bsi_{4\dot{1}}) \; , \nln
\Theta_4 & = -\ihalf (\Psi_{1\dot{3}} - \Psi_{2\dot{4}}) \; , &
\Theta_8 & = -\ihalf (\Bsi_{3\dot{1}} - \Bsi_{4\dot{2}}) \;   \nn
\end{align}
and $\Theta_{9\ldots16} = 0$. The fermions in the original action of Maldacena and Swanson \cite{Maldacena:2006rv} are obtained by splitting $\Theta$ into real and imaginary part as follows
\[ \label{eqn:MS-fermion}
 \Theta = \Pi \psi_+ + i \psi_- \; .
\]
The free action for the fermions in the two different notations read
\be \label{eqn:free-fermion-action}
\Lagr_0^{\mathrm{fermions}}
 \eq   2i \psi_- \partial_+ \psi_-
     + 2i \psi_+ \partial_- \psi_+
     + 2im \psi_- \Pi \psi_+ \\
 \eq + \tfrac{i}{2} \Psi^*_{\lAA\raa} \tim{\Psi}^{\lAA\raa}
     - \tfrac{i}{4} \brk{\Psi^*_{\lAA\raa} \spa{\Psi}^{*\lAA\raa}
                       + \Psi  _{\lAA\raa} \spa{\Psi}^{ \lAA\raa} }
     - \tfrac{m}{2} \, \Psi^*_{\lAA\raa} \Psi^{\lAA\raa} \nl
     + \tfrac{i}{2} \Bsi^*_{\laa\rAA} \tim{\Bsi}^{\laa\rAA}
     - \tfrac{i}{4} \brk{\Bsi^*_{\laa\rAA} \spa{\Bsi}^{*\laa\rAA}
                       + \Bsi  _{\laa\rAA} \spa{\Bsi}^{ \laa\rAA} }
     - \tfrac{m}{2} \, \Bsi^*_{\laa\rAA} \Bsi^{\laa\rAA} \nn \; .
\ee
Introducing the fermions $\psi_\pm$ is necessary for taking the near-flat-space limit in which the plus and minus components are treated differently. However, having arrived in near-flat space this split is not convenient any more. This is because of the mass term in \eqref{eqn:free-fermion-action} which couples $\psi_+$ and $\psi_-$ with the consequence that both fields are part of the same physical excitation. Therefore it is useful to re-combine them. One way would be to form a two-component world-sheet spinor. However, since in the special case at hand $\psi_+$ enters only the quadratic action, it is easiest to eliminate $\psi_+$ by integrating it out. The remaining fermion is then simply denoted by $\psi$ without subscript and the free action becomes
\[
\Lagr_0^{\mathrm{fermions}} = \ihalf \, \psi \, \frac{\partial^2+m^2}{\partial_-} \, \psi
\]
and the mapping \eqref{eqn:MS-fermion} turns into
\[
 \Theta = \lrbrk{ i - \frac{2\partial_+}{m} } \, \psi
        = \lrbrk{ i + \frac{m}{2\partial_-} } \, \psi \; .
\]
In conclusion, the identification of the corresponding states is
\begin{align}
\ket{\psi_1} & = + \half (\ket{\Psi_{1\dot{3}}} + \ket{\Psi_{2\dot{4}}}) \; , &
\ket{\psi_5} & = + \half (\ket{\Bsi_{3\dot{1}}} + \ket{\Bsi_{4\dot{2}}}) \; , \nln
\ket{\psi_2} & = -\ihalf (\ket{\Psi_{1\dot{4}}} + \ket{\Psi_{2\dot{3}}}) \; , &
\ket{\psi_6} & = -\ihalf (\ket{\Bsi_{3\dot{2}}} + \ket{\Bsi_{4\dot{1}}}) \; , \label{eqn:su2-ferm-ket} \\
\ket{\psi_3} & = + \half (\ket{\Psi_{1\dot{4}}} - \ket{\Psi_{2\dot{3}}}) \; , &
\ket{\psi_7} & = + \half (\ket{\Bsi_{3\dot{2}}} - \ket{\Bsi_{4\dot{1}}}) \; , \nln
\ket{\psi_4} & = -\ihalf (\ket{\Psi_{1\dot{3}}} - \ket{\Psi_{2\dot{4}}}) \; , &
\ket{\psi_8} & = -\ihalf (\ket{\Bsi_{3\dot{1}}} - \ket{\Bsi_{4\dot{2}}}) \; . \nn
\end{align}

%%%%%%%%%%%%%%%%%%%%%%%%%%%%%%%%%%%%%%%%%%%%%%%%%%%%%%%%%%%%%%%%%%%%%%%%%%%
%%%%%%%%%%%%%%%%%%%%%%%%%%%%%%%%%%%%%%%%%%%%%%%%%%%%%%%%%%%%%%%%%%%%%%%%%%%
\section{Phase space} \label{app:phasespace}

\begin{figure}
\begin{center}
\includegraphics[scale=0.5]{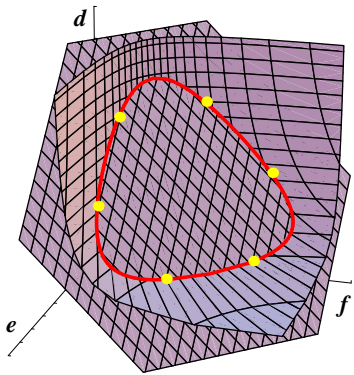}
\end{center}
\caption{\textbf{Three particle phase space.} Energy and momentum conservation forces the set of light-cone momenta $\{d,e,f\}$ to lie on a curve given by the intersection of a plane with a 3d-hyperbola. The consequence of the higher charges is to fix the momenta to permutations of the initial momenta, which is indicated by the dots on the curve.}
\label{fig:phasespace}
\end{figure}

Factorized scattering implies that the set of initial momenta is equal to the set of final momenta. This is a consequence of the existence of infinitely many conserved charges in integrable field theories. This makes the phase space trivial.

However, in perturbative computations using computing Feynman diagrams, only energy and momentum conservation are explicit. Hence, the three-particle phase space is a priori one-dimensional. In terms of light-cone momenta $d$, $e$ and $f$ it is given as the solution of
\[ \label{eqn:phasespace}
  d + e + f = A
  \comma
  \frac{1}{d} + \frac{1}{e} + \frac{1}{f} = \frac{3}{B} \; .
\]
We use $d$, $e$, $f$ for the particles in the final state. The constants $A$ and $B$ characterize the total energy and momentum, computed from the initial momenta $a$, $b$ and $c$ through corresponding formulas.

The surfaces defined by the equations \eqref{eqn:phasespace} are drawn in \figref{fig:phasespace}. The phase space, geometrically given by their intersection, is a planar elliptic curve of degree three. This can be seen after the coordinate transformation
\[
  x = -d-e+2f
  \comma
  y = -d+2e-f
  \comma
  z = d+e+f
\]
producing
\[
  x^2 y + x y^2 + (A-B)(x^2+y^2+x y) - A^2(A-3B) = 0
  \comma z = A \; .
\]

The scattering amplitudes computed in the main text are evaluated on the entire one-dimensional phase space. However, it turns out that the amplitudes vanish outside the six special points, which are located there where the final momenta are pairwise equal to the initial momenta. If expressions for the higher conserved charges were know, one could include corresponding surfaces in the figure \figref{fig:phasespace} and observe that the intersections of all surfaces are precisely these six points.

%%%%%%%%%%%%%%%%%%%%%%%%%%%%%%%%%%%%%%%%%%%%%%%%%%%%%%%%%%%%%%%%%%%%%%%%%%%
%%%%%%%%%%%%%%%%%%%%%%%%%%%%%%%%%%%%%%%%%%%%%%%%%%%%%%%%%%%%%%%%%%%%%%%%%%%
\section{Feynman diagrams and integrals} \label{app:diagrams}

%%%%%%%%%%%%%%%%%%%%%%%%%%%%%%%%%%%%%%%%%%%%%%%%%%%%%%%%%%%%%%%%%%%%%%%%%%%%%%%%
\paragraph{Tree-level.} The tree-level diagrams are essentially given by a single propagator. As sketched in \figref{fig:tree}, there are three momenta flowing into either side and $r$ powers of internal momentum:

\[ \label{eqn:tree}
 \prop_r(\pin) = \frac{(a+b+c)^r \, (2\pi)^2 \delta^2(\vecpin)}{(\vec{a}+\vec{b}+\vec{c})^2 - m^2 + i \eps} \; .
\]
Obviously
\[ \label{eqn:tree-relation}
 \prop_r(\pin) = (a+b+c)^r \prop_0(\pin) \; .
\]

%%%%%%%%%%%%%%%%%%%%%%%%%%%%%%%%%%%%%%%%%%%%%%%%%%%%%%%%%%%%%%%%%%%%%%%%%%%%%%%%
\paragraph{Bubble integral.} We compute the single bubble integral, cf. \figref{fig:bubble}. With $r$ and $s$ powers of momentum inserted, the integral reads
\[ \label{eqn:bubble}
  \bubble_{rs}(a,b) = \int \frac{d^2\vec{k}}{(2\pi)^2} \frac{k^r \, (a+b-k)^s }{[\vec{k}^2 - m^2 + i\eps][(\vec{a}+\vec{b}-\vec{k})^2 - m^2 + i\eps]} \; .
\]
Note the symmetry $\bubble_{rs} = \bubble_{sr}$. Bubbles with different numbers of momentum insertions are related by
\[ \label{eqn:bubble-relation}
  \bubble_{r+1,s}(a,b) + \bubble_{r,s+1}(a,b) = (a+b) \bubble_{rs}(a,b) \; .
\]
By means of this relation, all expressions occurring in our computations can be simplified such that we only need to explicitly evaluate the integral for $r=s=0$. In this case one finds
\[ \label{oneloopbubble00}
\bubble_{00}(a,b) = \frac{i}{2\pi m^2} \, \frac{a \, b}{a^2 - b^2} \ln\lrbrk{-\frac{b}{a}} \; ,
\]
where for $-\frac{b}{a} < 0$ the logarithm is to be understood as
\[
  \ln\lrbrk{-\frac{b}{a}} = -i\pi \, \sign\lrbrk{\frac{b}{a}-1} + \ln\frac{b}{a} \; .
\]
This way of analytical continuation is dictated by the $+i\eps$-prescription in \eqref{eqn:bubble}.

%%%%%%%%%%%%%%%%%%%%%%%%%%%%%%%%%%%%%%%%%%%%%%%%%%%%%%%%%%%%%%%%%%%%%%%%%%%%%%%%
\paragraph{Dog diagram.} The first kind of one-loop diagrams \figref{fig:dog}, remotely resembling a dog, is nothing but a product of the bubble \figref{fig:bubble} and a tree diagram \figref{fig:tree}
\[
  \dog_{rst}(\pin) = \bubble_{rs}(a,b) \, \prop_{t}(\pin) \; .
\]
The identities for the bubble \eqref{eqn:bubble-relation} and the propagator \eqref{eqn:tree-relation}, respectively, immediately imply
\[ \label{eqn:dog-relation}
  \dog_{r+1,s,t}(\pin) + \dog_{r,s+1,t}(\pin) = (a+b) \, \dog_{rst}(\pin)
  \comma
  \dog_{r,s,t+1}(\pin) = (a+b+c) \, \dog_{rst}(\pin) \; .
\]

%%%%%%%%%%%%%%%%%%%%%%%%%%%%%%%%%%%%%%%%%%%%%%%%%%%%%%%%%%%%%%%%%%%%%%%%%%%%%%%%
\paragraph{Sun integral.} To the second kind of one-loop diagrams \figref{fig:sun} we refer to as sun. The corresponding integral is given by
\[ \label{eqn:sun}
  \sun_{rst}(\eta) = \int \frac{d^2\vec{k}}{(2\pi)^2} \frac{k^s \, (k+a+b)^t \, (k+a+b+c+d)^r \, (2\pi)^2 \delta^2(\vec{a}+\vec{b}+\vec{c}+\vec{d}+\vec{e}+\vec{f})}{[\vec{k}^2 - m^2][(\vec{k}+\vec{a}+\vec{b})^2 - m^2][(\vec{k}+\vec{a}+\vec{b}+\vec{c}+\vec{d})^2 - m^2]} \; ,
\]
where the $+i\eps$ pole prescription is understood. The integral is symmetric under even permutations of the triples $(r,a,b)$, $(s,c,d)$ and $(t,e,f)$, and it acquires a sign $(-)^{r+s+t}$ under odd permutations, e.g. $\sun_{rst}(a,b,c,d,e,f) = (-)^{r+s+t} \sun_{srt}(c,d,a,b,e,f)$. Furthermore, integrals with different numbers of derivatives are related by the following formulas
\begin{align}
  \sun_{r,s,t+1}(\pin) - \sun_{r,s+1,t}(\pin) & = (a+b) \, \sun_{rst}(\pin) \; , \nn \\
  \sun_{r+1,s,t}(\pin) - \sun_{r,s,t+1}(\pin) & = (c+d) \, \sun_{rst}(\pin) \; , \label{eqn:sun-relation} \\
  \sun_{r,s+1,t}(\pin) - \sun_{r+1,s,t}(\pin) & = (e+f) \, \sun_{rst}(\pin) \; . \nn
\end{align}

The evaluation of the integral \eqref{eqn:sun} is most conveniently done in the light-cone coordinates. After integrating the plus momentum, we use a partial fractions expansion \cite{Kallen:1965} to compute the remaining integral over the minus component. In this way it is possible to see how the sun is reduced to a sum of bubble integrals each multiplied by a tree-level diagram $I_r$, in agreement with the cutting rules for one-loop diagrams in two dimensions \cite{Kallen:1965}.
In the case $r+s+t \le 2$, which is relevant to our computations, the result is given by
\be \label{eqn:cutting-rule}
  \sun_{rst}(\pin) \eq
  + \half \bigsbrk{ a^t \, (-b)^s \, I_r(a,c,d,b,e,f) + b^t \, (-a)^s \, I_r(a,c,d,b,e,f) } \bubble_{00}(a,b) \nl
  + \half \bigsbrk{ c^r \, (-d)^t \, I_s(c,e,f,d,a,b) + d^r \, (-c)^t \, I_s(d,e,f,c,a,b) } \bubble_{00}(c,d) \nl
  + \half \bigsbrk{ e^s \, (-f)^r \, I_t(e,a,b,f,c,d) + f^s \, (-e)^r \, I_t(f,a,b,e,c,d) } \bubble_{00}(e,f) \; .
\ee
The pole prescription for the propagator in the tree level part turns out to be momentum dependent: the sign of $i\eps$ in $I_r(a,b,c,d,e,f)$ is equal to the
\[ \label{eqn:momentum-dependent-eps}
\sign \frac{(\vec{d}-\vec{a})\cdot(\vec{b}+\vec{c})}{(\vec{d}-\vec{a})^2} \; .
\]
This non-trivial prescription is essential for correctly extracting the poles in the three-particle S-matrix.

%%%%%%%%%%%%%%%%%%%%%%%%%%%%%%%%%%%%%%%%%%%%%%%%%%%%%%%%%%%%%%%%%%%%%%%%%%%
%%%%%%%%%%%%%%%%%%%%%%%%%%%%%%%%%%%%%%%%%%%%%%%%%%%%%%%%%%%%%%%%%%%%%%%%%%%
\section{Some scattering amplitudes}
\label{app:scatt-amp}

To facilitate the readability of the main text, we have moved some of the amplitudes discussed in \secref{sec:bonus-processes} into this appendix. The following amplitudes are given up to an additive non-singular function $\Ramp(\pin)$ which cancel in the sum $\Amp^{\mathrm{1-loop}}(\pin) = \Amp^{\mathrm{dog}}(\pin) + \Amp^{\mathrm{sun}}(\pin)$, cf. \eqref{eqn:oneloop-sample-dog} and \eqref{eqn:oneloop-sample-sun}.
\paragraph{$Y_1(a) \psi_1(b) \psi_6(c) \rightarrow Z_7(d) Y_2(e) Y_3(f)$}
\be
\label{eqn:mixed-dog}
\Amp^{\mathrm{dog}}(\pin) \eq
  + \frac{i\g^3a^{3}b^{3/2}c^{3/2}\bigsbrk{(a-b)(b-c)c^2+a^2(b+c)^2}}{(b-a)(c^2-a^2)(b+c)} \, \bigsbrk{
       \delta_{ad} \delta_{be} \delta_{cf}
     - \delta_{ad} \delta_{ce} \delta_{bf} } \nl[2mm]
  - \frac{i\g^3a^3b^{3/2}c^{3/2}\bigsbrk{c^2(a^2-b^2)+a^2(b(b+a)+c(b-a))}}{(c^2-a^2)(b^2-a^2)} \nl[1mm]
    \quad \times \bigsbrk{
       \delta_{bd} \delta_{ae} \delta_{cf}
     - \delta_{cd} \delta_{be} \delta_{af}
     + \delta_{cd} \delta_{ae} \delta_{bf}
     - \delta_{bd} \delta_{ce} \delta_{af} }
\\[6mm]
\label{eqn:mixed-sun}
\Amp^{\mathrm{sun}}(\pin) \eq
  + \frac{i\g^3a^{3}b^{5/2}c^{5/2}(b-c)}{(b-a)(c-a)(b+c)} \, \bigsbrk{
          \delta_{ad} \delta_{be} \delta_{cf}
        - \delta_{ad} \delta_{ce} \delta_{bf}} \nl[2mm]
  - \frac{i\g^3 a^3 b^{5/2}c^{5/2}[c(b+a)+a(b-a)]}{(b^2-a^2)(c^2-a^2)} \, \bigsbrk{
          \delta_{bd} \delta_{ae} \delta_{cf}
        - \delta_{bd} \delta_{ce} \delta_{af}} \nl[2mm]
  - \frac{i\g^3a^3b^{5/2}c^{5/2}[c(b-a)-a(b+a)]}{(b^2-a^2)(c^2-a^2)} \, \bigsbrk{
          \delta_{cd} \delta_{ae} \delta_{bf}
       -  \delta_{cd} \delta_{be} \delta_{af}}
\ee
\paragraph{$Z_7(a) Y_2(b) Y_3(c) \rightarrow Y_1(d) \psi_1(e) \psi_6(f)$}
\begin{align}
\Amp^{\mathrm{dog}}(\pin)
  & = \frac{i\g^3a^3b^{3/2}c^{3/2}\bigsbrk{a^2(b+c)^2+c^2(c-b)(b-a)}}{(a-b)(a^2-c^2)(b+c)} \,\bigsbrk{
                           \delta_{ad} \delta_{be} \delta_{cf}
                         - \delta_{ad} \delta_{ce} \delta_{bf}} \\[2mm]
  & - \frac{i\g^3a^{3/2}c^{3/2}b^2[2ab^2c^3-b^3c^2(b+c)+a^3(b^2-c^2)(b+2c)+a^2b(b^3-2bc^2+c^3)]}{(b^2-a^2)(c-a)(c^2-b^2)} \nn \\[1mm]
  &   \quad\times \bigsbrk{\delta_{bd} \delta_{ae} \delta_{cf}
                         - \delta_{bd} \delta_{af} \delta_{ce}} \nn \\[4mm]
  & + \frac{i\g^3a^{3/2}b^{3/2}c^3\bigsbrk{b^2(a^2-c^2)+c^2(a(c-a)-b(a+c))}}{(c^2-a^2)(c^2-b^2)} \, \bigsbrk{
                           \delta_{cd} \delta_{be} \delta_{af}
                         - \delta_{cd} \delta_{bf} \delta_{ae}} \nn
\\[6mm]
\Amp^{\mathrm{sun}}(\pin)
  & = \frac{i\g^3a^3b^{5/2}c^{5/2}(c-b)}{(b-a)(c-a)(b+c)} \, \bigsbrk{
                           \delta_{ad} \delta_{ce} \delta_{bf}
                         - \delta_{ad} \delta_{be} \delta_{cf}} \\[2mm]
  & + \frac{i\g^3b^3a^{5/2}c^{5/2}[a(c-b)(c^2-a^2)+a^2(b+c)(3b-4c)+bc^2(b+c)]}{(b^2-a^2)(c^2-a^2)(c^2-b^2)} \nn \\[1mm]
  &   \quad\times \bigsbrk{\delta_{bd} \delta_{ce} \delta_{af}
                         - \delta_{bd} \delta_{ae} \delta_{cf}} \nn \\[4mm]
  & + \frac{i\g^3c^3a^{5/2}b^{5/2}[a(c-b)+c(c+b)]}{(c^2-a^2)(c^2-b^2)} \, \bigsbrk{
                           \delta_{cd} \delta_{be} \delta_{af}
                         - \delta_{cd} \delta_{bf} \delta_{ae}} \nn
\end{align}

%%%%%%%%%%%%%%%%%%%%%%%%%%%%%%%%%%%%%%%%%%%%%%%%%%%%%%%%%%%%%%%%%%%%%%%%%%%
\bibliographystyle{nb}
\bibliography{factscatt}

\begin{thebibliography}{10}
\ifx\href\asklfhas\newcommand{\href}[2]{#2}\fi
\raggedright
\small
\parskip 0pt

%%CITATION = HEP-TH/0206103;%%
\bibitem{Mandal:2002fs}
G.~Mandal, N.~V.~Suryanarayana and S.~R.~Wadia,
\textit{``Aspects of semiclassical strings in $AdS_5$''},
\textsf{Phys.~Lett.~B543,~81~(2002)},
\href{http://arXiv.org/abs/hep-th/0206103}{\texttt{hep-th/0206103}}.
%
%%CITATION = HEP-TH 0305116;%%
\bibitem{Bena:2003wd}
I.~Bena, J.~Polchinski and R.~Roiban,
\textit{``Hidden symmetries of the $AdS_5\times S^5$ superstring''},
\textsf{Phys.~Rev.~D69,~046002~(2004)},
\href{http://arXiv.org/abs/hep-th/0305116}{\texttt{hep-th/0305116}}.
%
%%CITATION = HEP-TH 0402207;%%
\bibitem{Kazakov:2004qf}
V.~A.~Kazakov, A.~Marshakov, J.~A.~Minahan and K.~Zarembo,
\textit{``Classical/quantum integrability in AdS/CFT''},
\textsf{JHEP~0405,~024~(2004)},
\href{http://arXiv.org/abs/hep-th/0402207}{\texttt{hep-th/0402207}}.
%
%%CITATION = HEP-TH 0212208;%%
\bibitem{Minahan:2002ve}
J.~A.~Minahan and K.~Zarembo,
\textit{``The Bethe-ansatz for $\mathcal{N}=4$ super Yang-Mills''},
\textsf{JHEP~0303,~013~(2003)},
\href{http://arXiv.org/abs/hep-th/0212208}{\texttt{hep-th/0212208}}.
%
%%CITATION = HEP-TH 0303060;%%
\bibitem{Beisert:2003tq}
N.~Beisert, C.~Kristjansen and M.~Staudacher,
\textit{``The dilatation operator of $\mathcal{N}=4$ conformal super Yang-Mills
  theory''},
\textsf{Nucl.~Phys.~B664,~131~(2003)},
\href{http://arXiv.org/abs/hep-th/0303060}{\texttt{hep-th/0303060}}.
%
%%CITATION = HEP-TH 0307042;%%
\bibitem{Beisert:2003yb}
N.~Beisert and M.~Staudacher,
\textit{``The $\mathcal{N}=4$ SYM Integrable Super Spin Chain''},
\textsf{Nucl.~Phys.~B670,~439~(2003)},
\href{http://arXiv.org/abs/hep-th/0307042}{\texttt{hep-th/0307042}}.
%
%%CITATION = HEP-TH 9711200;%%
\bibitem{Maldacena:1998re}
J.~M.~Maldacena,
\textit{``The large N limit of superconformal field theories and
  supergravity''},
\textsf{Adv.~Theor.~Math.~Phys.~2,~231~(1998)},
\href{http://arXiv.org/abs/hep-th/9711200}{\texttt{hep-th/9711200}}.
%
%%CITATION = HEP-TH 9802109;%%
\bibitem{Gubser:1998bc}
S.~S.~Gubser, I.~R.~Klebanov and A.~M.~Polyakov,
\textit{``Gauge theory correlators from non-critical string theory''},
\textsf{Phys.~Lett.~B428,~105~(1998)},
\href{http://arXiv.org/abs/hep-th/9802109}{\texttt{hep-th/9802109}}.
%
%%CITATION = HEP-TH 9802150;%%
\bibitem{Witten:1998qj}
E.~Witten,
\textit{``Anti-de Sitter space and holography''},
\textsf{Adv.~Theor.~Math.~Phys.~2,~253~(1998)},
\href{http://arXiv.org/abs/hep-th/9802150}{\texttt{hep-th/9802150}}.
%
%%CITATION = HEP-TH/0407277;%%
\bibitem{Beisert:2004ry}
N.~Beisert,
\textit{``The dilatation operator of $\mathcal{N} = 4$ super Yang-Mills theory
  and integrability''},
\textsf{Phys.~Rept.~405,~1~(2005)},
\href{http://arXiv.org/abs/hep-th/0407277}{\texttt{hep-th/0407277}}.
%
%%CITATION = HEP-TH/0411191;%%
\bibitem{Zarembo:2004hp}
K.~Zarembo,
\textit{``Semiclassical Bethe ansatz and AdS/CFT''},
\textsf{Comptes~Rendus~Physique~5,~1081~(2004)},
\href{http://arXiv.org/abs/hep-th/0411191}{\texttt{hep-th/0411191}}.
%
%%CITATION = HEP-TH/0507136;%%
\bibitem{Plefka:2005bk}
J.~Plefka,
\textit{``Spinning strings and integrable spin chains in the AdS/CFT
  correspondence''},
\href{http://arXiv.org/abs/hep-th/0507136}{\texttt{hep-th/0507136}}.
%
%%CITATION = JPAGB,A39,12657;%%
\bibitem{Minahan:2006sk}
J.~A.~Minahan,
\textit{``A brief introduction to the Bethe ansatz in $\mathcal{N}=4$
  super-Yang- Mills''},
\textsf{J.~Phys.~A39,~12657~(2006)}.
%
%%CITATION = HEP-TH 0504190;%%
\bibitem{Beisert:2005fw}
N.~Beisert and M.~Staudacher,
\textit{``Long-range $PSU(2,2|4)$ Bethe ansaetze for gauge theory and
  strings''},
\textsf{Nucl.~Phys.~B727,~1~(2005)},
\href{http://arXiv.org/abs/hep-th/0504190}{\texttt{hep-th/0504190}}.
%
%%CITATION = HEP-TH 0412188;%%
\bibitem{Staudacher:2004tk}
M.~Staudacher,
\textit{``The factorized S-matrix of CFT/AdS''},
\textsf{JHEP~0505,~054~(2005)},
\href{http://arXiv.org/abs/hep-th/0412188}{\texttt{hep-th/0412188}}.
%
%%CITATION = HEP-TH 0511082;%%
\bibitem{Beisert:2005tm}
N.~Beisert,
\textit{``The $su(2|2)$ dynamic S-matrix''},
\href{http://arXiv.org/abs/hep-th/0511082}{\texttt{hep-th/0511082}}.
%
%%CITATION = HEP-TH/0612229;%%
\bibitem{Arutyunov:2006yd}
G.~Arutyunov, S.~Frolov and M.~Zamaklar,
\textit{``The Zamolodchikov-Faddeev algebra for $AdS_5 \times S^5$
  superstring''},
\textsf{JHEP~0704,~002~(2007)},
\href{http://arXiv.org/abs/hep-th/0612229}{\texttt{hep-th/0612229}}.
%
%%CITATION = HEP-TH 0603038;%%
\bibitem{Janik:2006dc}
R.~A.~Janik,
\textit{``The $AdS_5 \times S^5$ superstring worldsheet S-matrix and crossing
  symmetry''},
\textsf{Phys.~Rev.~D73,~086006~(2006)},
\href{http://arXiv.org/abs/hep-th/0603038}{\texttt{hep-th/0603038}}.
%
%%CITATION = HEP-TH 0406256;%%
\bibitem{Arutyunov:2004vx}
G.~Arutyunov, S.~Frolov and M.~Staudacher,
\textit{``Bethe ansatz for quantum strings''},
\textsf{JHEP~0410,~016~(2004)},
\href{http://arXiv.org/abs/hep-th/0406256}{\texttt{hep-th/0406256}}.
%
%%CITATION = HEP-TH 0603204;%%
\bibitem{Hernandez:2006tk}
R.~Hernandez and E.~Lopez,
\textit{``Quantum corrections to the string Bethe ansatz''},
\textsf{JHEP~0607,~004~(2006)},
\href{http://arXiv.org/abs/hep-th/0603204}{\texttt{hep-th/0603204}}.
%
%%CITATION = HEP-TH 0609044;%%
\bibitem{Beisert:2006ib}
N.~Beisert, R.~Hernandez and E.~Lopez,
\textit{``A crossing-symmetric phase for $AdS_5 \times S^5$ strings''},
\textsf{JHEP~0611,~070~(2006)},
\href{http://arXiv.org/abs/hep-th/0609044}{\texttt{hep-th/0609044}}.
%
%%CITATION = HEP-TH/0610251;%%
\bibitem{Beisert:2006ez}
N.~Beisert, B.~Eden and M.~Staudacher,
\textit{``Transcendentality and crossing''},
\textsf{J.~Stat.~Mech.~0701,~P021~(2007)},
\href{http://arXiv.org/abs/hep-th/0610251}{\texttt{hep-th/0610251}}.
%
%%CITATION = TMPHA,26,132;%%
\bibitem{Kulish:1975ba}
P.~P.~Kulish,
\textit{``Factorization of the classical and quantum S-matrix and conservation
  laws''},
\textsf{Theor.~Math.~Phys.~26,~132~(1976)}.
%
%%CITATION = NUPHA,B135,1;%%
\bibitem{Luscher:1977uq}
M.~L{\"u}scher,
\textit{``Quantum Nonlocal Charges and Absence of Particle Production in the
  Two-Dimensional Nonlinear Sigma Model''},
\textsf{Nucl.~Phys.~B135,~1~(1978)}.
%
%%CITATION = PHRVA,D17,2134;%%
\bibitem{Shankar:1977cm}
R.~Shankar and E.~Witten,
\textit{``The S-matrix of the supersymmetric nonlinear sigma model''},
\textsf{Phys.~Rev.~D17,~2134~(1978)}.
%
%%CITATION = APNYA,120,253;%%
\bibitem{Zamolodchikov:1978xm}
A.~B.~Zamolodchikov and A.~B.~Zamolodchikov,
\textit{``Factorized S-matrices in two dimensions as the exact solutions of
  certain relativistic quantum field models''},
\textsf{Annals~Phys.~120,~253~(1979)}.
%
%%CITATION = HEP-TH/9810026;%%
\bibitem{Dorey:1996gd}
P.~Dorey,
\textit{``Exact S-matrices''},
\href{http://arXiv.org/abs/hep-th/9810026}{\texttt{hep-th/9810026}}.
%
%%CITATION = HEP-TH/0610248;%%
\bibitem{Bern:2006ew}
Z.~Bern, M.~Czakon, L.~J.~Dixon, D.~A.~Kosower and V.~A.~Smirnov,
\textit{``The Four-Loop Planar Amplitude and Cusp Anomalous Dimension in
  Maximally Supersymmetric Yang-Mills Theory''},
\textsf{Phys.~Rev.~D75,~085010~(2007)},
\href{http://arXiv.org/abs/hep-th/0610248}{\texttt{hep-th/0610248}}.
%
%%CITATION = HEP-TH/0612309;%%
\bibitem{Cachazo:2006az}
F.~Cachazo, M.~Spradlin and A.~Volovich,
\textit{``Four-Loop Cusp Anomalous Dimension From Obstructions''},
\textsf{Phys.~Rev.~D75,~105011~(2007)},
\href{http://arXiv.org/abs/hep-th/0612309}{\texttt{hep-th/0612309}}.
%
%%CITATION = HEP-TH 0611135;%%
\bibitem{Benna:2006nd}
M.~K.~Benna, S.~Benvenuti, I.~R.~Klebanov and A.~Scardicchio,
\textit{``A test of the AdS/CFT correspondence using high-spin operators''},
\href{http://arXiv.org/abs/hep-th/0611135}{\texttt{hep-th/0611135}}.
%
%%CITATION = HEP-TH/0702028;%%
\bibitem{Alday:2007qf}
L.~F.~Alday, G.~Arutyunov, M.~K.~Benna, B.~Eden and I.~R.~Klebanov,
\textit{``On the strong coupling scaling dimension of high spin operators''},
\textsf{JHEP~0704,~082~(2007)},
\href{http://arXiv.org/abs/hep-th/0702028}{\texttt{hep-th/0702028}}.
%
%%CITATION = HEP-TH/0703031;%%
\bibitem{Kostov:2007kx}
I.~Kostov, D.~Serban and D.~Volin,
\textit{``Strong coupling limit of Bethe ansatz equations''},
\href{http://arXiv.org/abs/hep-th/0703031}{\texttt{hep-th/0703031}}.
%
%%CITATION = HEP-TH/0703131;%%
\bibitem{Beccaria:2007tk}
M.~Beccaria, G.~F.~De~Angelis and V.~Forini,
\textit{``The scaling function at strong coupling from the quantum string Bethe
  equations''},
\textsf{JHEP~0704,~066~(2007)},
\href{http://arXiv.org/abs/hep-th/0703131}{\texttt{hep-th/0703131}}.
%
%%CITATION = ARXIV:0705.0890;%%
\bibitem{Casteill:2007ct}
P.~Y.~Casteill and C.~Kristjansen,
\textit{``The Strong Coupling Limit of the Scaling Function from the Quantum
  String Bethe Ansatz''},
\href{http://arXiv.org/abs/arXiv:0705.0890 [hep-th]}{\texttt{arXiv:0705.0890
  [hep-th]}}.
%
%%CITATION = ARXIV:0704.3638;%%
\bibitem{Roiban:2007jf}
R.~Roiban, A.~Tirziu and A.~A.~Tseytlin,
\textit{``Two-loop world-sheet corrections in $AdS_5 \times S^5$
  superstring''},
\href{http://arXiv.org/abs/arXiv:0704.3638 [hep-th]}{\texttt{arXiv:0704.3638
  [hep-th]}}.
%
%%CITATION = ARXIV:0705.0321;%%
\bibitem{Beisert:2007hz}
N.~Beisert, T.~McLoughlin and R.~Roiban,
\textit{``The Four-Loop Dressing Phase of $\mathcal{N} = 4$ SYM''},
\href{http://arXiv.org/abs/arXiv:0705.0321 [hep-th]}{\texttt{arXiv:0705.0321
  [hep-th]}}.
%
%%CITATION = HEP-TH/0701240;%%
\bibitem{Klose:2007wq}
T.~Klose and K.~Zarembo,
\textit{``Reduced sigma-model on $AdS_5 \times S^5$: one-loop scattering
  amplitudes''},
\textsf{JHEP~0702,~071~(2007)},
\href{http://arXiv.org/abs/hep-th/0701240}{\texttt{hep-th/0701240}}.
%
%%CITATION = ARXIV:0704.3891;%%
\bibitem{Klose:2007rz}
T.~Klose, T.~McLoughlin, J.~A.~Minahan and K.~Zarembo,
\textit{``World-sheet scattering in $AdS_5 \times S^5$ at two loops''},
\href{http://arXiv.org/abs/arXiv:0704.3891 [hep-th]}{\texttt{arXiv:0704.3891
  [hep-th]}}.
%
%%CITATION = ARXIV:0707.0668;%%
\bibitem{Chen:2007vs}
H.-Y.~Chen, N.~Dorey and R.~F.~L.~Matos,
\textit{``Quantum Scattering of Giant Magnons''},
\href{http://arXiv.org/abs/arXiv:0707.0668 [hep-th]}{\texttt{arXiv:0707.0668
  [hep-th]}}.
%
%%CITATION = HEP-TH 0204226;%%
\bibitem{Frolov:2002av}
S.~Frolov and A.~A.~Tseytlin,
\textit{``Semiclassical quantization of rotating superstring in $AdS_5 \times
  S^5$''},
\textsf{JHEP~0206,~007~(2002)},
\href{http://arXiv.org/abs/hep-th/0204226}{\texttt{hep-th/0204226}}.
%
%%CITATION = HEP-TH 0306130;%%
\bibitem{Frolov:2003tu}
S.~Frolov and A.~A.~Tseytlin,
\textit{``Quantizing three-spin string solution in $AdS_5 \times S^5$''},
\textsf{JHEP~0307,~016~(2003)},
\href{http://arXiv.org/abs/hep-th/0306130}{\texttt{hep-th/0306130}}.
%
%%CITATION = HEP-TH 0408187;%%
\bibitem{Frolov:2004bh}
S.~A.~Frolov, I.~Y.~Park and A.~A.~Tseytlin,
\textit{``On one-loop correction to energy of spinning strings in $S^5$''},
\textsf{Phys.~Rev.~D71,~026006~(2005)},
\href{http://arXiv.org/abs/hep-th/0408187}{\texttt{hep-th/0408187}}.
%
%%CITATION = HEP-TH 0501203;%%
\bibitem{Park:2005ji}
I.~Y.~Park, A.~Tirziu and A.~A.~Tseytlin,
\textit{``Spinning strings in $AdS_5 \times S^5$: One-loop correction to energy
  in $SL(2)$ sector''},
\textsf{JHEP~0503,~013~(2005)},
\href{http://arXiv.org/abs/hep-th/0501203}{\texttt{hep-th/0501203}}.
%
%%CITATION = HEP-TH/0611269;%%
\bibitem{Frolov:2006qe}
S.~Frolov, A.~Tirziu and A.~A.~Tseytlin,
\textit{``Logarithmic corrections to higher twist scaling at strong coupling
  from AdS/CFT''},
\textsf{Nucl.~Phys.~B766,~232~(2007)},
\href{http://arXiv.org/abs/hep-th/0611269}{\texttt{hep-th/0611269}}.
%
%%CITATION = HEP-TH 0507189;%%
\bibitem{Schafer-Nameki:2005tn}
S.~Schafer-Nameki, M.~Zamaklar and K.~Zarembo,
\textit{``Quantum corrections to spinning strings in $AdS_5 \times S^5$ and
  Bethe ansatz: A comparative study''},
\textsf{JHEP~0509,~051~(2005)},
\href{http://arXiv.org/abs/hep-th/0507189}{\texttt{hep-th/0507189}}.
%
%%CITATION = HEP-TH 0509084;%%
\bibitem{Beisert:2005cw}
N.~Beisert and A.~A.~Tseytlin,
\textit{``On quantum corrections to spinning strings and Bethe equations''},
\textsf{Phys.~Lett.~B629,~102~(2005)},
\href{http://arXiv.org/abs/hep-th/0509084}{\texttt{hep-th/0509084}}.
%
%%CITATION = HEP-TH 0509096;%%
\bibitem{Schafer-Nameki:2005is}
S.~Schafer-Nameki and M.~Zamaklar,
\textit{``Stringy sums and corrections to the quantum string Bethe ansatz''},
\textsf{JHEP~0510,~044~(2005)},
\href{http://arXiv.org/abs/hep-th/0509096}{\texttt{hep-th/0509096}}.
%
%%CITATION = HEP-TH 0602214;%%
\bibitem{Schafer-Nameki:2006gk}
S.~Schafer-Nameki,
\textit{``Exact expressions for quantum corrections to spinning strings''},
\textsf{Phys.~Lett.~B639,~571~(2006)},
\href{http://arXiv.org/abs/hep-th/0602214}{\texttt{hep-th/0602214}}.
%
%%CITATION = HEP-TH 0604069;%%
\bibitem{Freyhult:2006vr}
L.~Freyhult and C.~Kristjansen,
\textit{``A universality test of the quantum string Bethe ansatz''},
\textsf{Phys.~Lett.~B638,~258~(2006)},
\href{http://arXiv.org/abs/hep-th/0604069}{\texttt{hep-th/0604069}}.
%
%%CITATION = HEP-TH 0610250;%%
\bibitem{Schafer-Nameki:2006ey}
S.~Schafer-Nameki, M.~Zamaklar and K.~Zarembo,
\textit{``How accurate is the quantum string Bethe ansatz?''},
\textsf{JHEP~0612,~020~(2006)},
\href{http://arXiv.org/abs/hep-th/0610250}{\texttt{hep-th/0610250}}.
%
%%CITATION = HEP-TH/0702151;%%
\bibitem{Rej:2007vm}
A.~Rej, M.~Staudacher and S.~Zieme,
\textit{``Nesting and dressing''},
\href{http://arXiv.org/abs/hep-th/0702151}{\texttt{hep-th/0702151}}.
%
%%CITATION = HEP-TH/0703177;%%
\bibitem{Sakai:2007rk}
K.~Sakai and Y.~Satoh,
\textit{``Origin of dressing phase in $\mathcal{N}=4$ Super Yang-Mills''},
\href{http://arXiv.org/abs/hep-th/0703177}{\texttt{hep-th/0703177}}.
%
%%CITATION = HEP-TH/0703266;%%
\bibitem{Gromov:2007cd}
N.~Gromov and P.~Vieira,
\textit{``Constructing the AdS/CFT dressing factor''},
\href{http://arXiv.org/abs/hep-th/0703266}{\texttt{hep-th/0703266}}.
%
%%CITATION = HEP-TH/0703104;%%
\bibitem{Dorey:2007xn}
N.~Dorey, D.~M.~Hofman and J.~Maldacena,
\textit{``On the singularities of the magnon S-matrix''},
\href{http://arXiv.org/abs/hep-th/0703104}{\texttt{hep-th/0703104}}.
%
%%CITATION = ARXIV:0707.0511;%%
\bibitem{Das:2007tb}
A.~Das, A.~Melikyan and V.~O.~Rivelles,
\textit{``The S-matrix of the Faddeev-Reshetikhin Model, Diagonalizability and
  PT Symmetry''},
\href{http://arXiv.org/abs/arXiv:0707.0511 [hep-th]}{\texttt{arXiv:0707.0511
  [hep-th]}}.
%
%%CITATION = HEP-TH/0310252;%%
\bibitem{Beisert:2003ys}
N.~Beisert,
\textit{``The $su(2|3)$ dynamic spin chain''},
\textsf{Nucl.~Phys.~B682,~487~(2004)},
\href{http://arXiv.org/abs/hep-th/0310252}{\texttt{hep-th/0310252}}.
%
%%CITATION = HEP-TH/0401057;%%
\bibitem{Serban:2004jf}
D.~Serban and M.~Staudacher,
\textit{``Planar N = 4 gauge theory and the Inozemtsev long range spin
  chain''},
\textsf{JHEP~0406,~001~(2004)},
\href{http://arXiv.org/abs/hep-th/0401057}{\texttt{hep-th/0401057}}.
%
%%CITATION = HEP-TH/0405001;%%
\bibitem{Beisert:2004hm}
N.~Beisert, V.~Dippel and M.~Staudacher,
\textit{``A novel long range spin chain and planar $\mathcal{N} = 4$ super
  Yang- Mills''},
\textsf{JHEP~0407,~075~(2004)},
\href{http://arXiv.org/abs/hep-th/0405001}{\texttt{hep-th/0405001}}.
%
%%CITATION = HEP-TH/0409180;%%
\bibitem{Agarwal:2004sz}
A.~Agarwal and S.~G.~Rajeev,
\textit{``Yangian symmetries of matrix models and spin chains: The dilatation
  operator of $\mathcal{N} = 4$ SYM''},
\textsf{Int.~J.~Mod.~Phys.~A20,~5453~(2005)},
\href{http://arXiv.org/abs/hep-th/0409180}{\texttt{hep-th/0409180}}.
%
%%CITATION = HEP-TH/0411170;%%
\bibitem{Berkovits:2004xu}
N.~Berkovits,
\textit{``Quantum consistency of the superstring in $AdS_5 \times S^5$
  background''},
\textsf{JHEP~0503,~041~(2005)},
\href{http://arXiv.org/abs/hep-th/0411170}{\texttt{hep-th/0411170}}.
%
%%CITATION = HEP-TH/0610283;%%
\bibitem{Zwiebel:2006cb}
B.~I.~Zwiebel,
\textit{``Yangian symmetry at two-loops for the $su(2|1)$ sector of
  $\mathcal{N}=4$ SYM''},
\textsf{J.~Phys.~A40,~1141~(2007)},
\href{http://arXiv.org/abs/hep-th/0610283}{\texttt{hep-th/0610283}}.
%
%%CITATION = ARXIV:0704.0400;%%
\bibitem{Beisert:2007ds}
N.~Beisert,
\textit{``The S-Matrix of AdS/CFT and Yangian Symmetry''},
\textsf{PoS~SOLVAY,~002~(2006)},
\href{http://arXiv.org/abs/arXiv:0704.0400 [nlin.SI]}{\texttt{arXiv:0704.0400
  [nlin.SI]}}.
%
%%CITATION = HEP-TH/0603157;%%
\bibitem{Eden:2006rx}
B.~Eden and M.~Staudacher,
\textit{``Integrability and transcendentality''},
\textsf{J.~Stat.~Mech.~0611,~P014~(2006)},
\href{http://arXiv.org/abs/hep-th/0603157}{\texttt{hep-th/0603157}}.
%
%%CITATION = HEP-TH/0405153;%%
\bibitem{Callan:2004ev}
C.~G.~Callan, T.~McLoughlin and I.~J.~Swanson,
\textit{``Higher impurity AdS/CFT correspondence in the near-BMN limit''},
\textsf{Nucl.~Phys.~B700,~271~(2004)},
\href{http://arXiv.org/abs/hep-th/0405153}{\texttt{hep-th/0405153}}.
%
%%CITATION = HEP-TH/0611169;%%
\bibitem{Klose:2006zd}
T.~Klose, T.~McLoughlin, R.~Roiban and K.~Zarembo,
\textit{``Worldsheet scattering in $AdS_5 \times S^5$''},
\textsf{JHEP~0703,~094~(2007)},
\href{http://arXiv.org/abs/hep-th/0611169}{\texttt{hep-th/0611169}}.
%
%%CITATION = HEP-TH/0703187;%%
\bibitem{Hentschel:2007xn}
A.~Hentschel, J.~Plefka and P.~Sundin,
\textit{``Testing the nested light-cone Bethe equations of the $AdS_5 \times
  S^5$ superstring''},
\textsf{JHEP~0705,~021~(2007)},
\href{http://arXiv.org/abs/hep-th/0703187}{\texttt{hep-th/0703187}}.
%
%%CITATION = ARXIV:0706.1525;%%
\bibitem{Mikhailov:2007mr}
A.~Mikhailov and S.~Schafer-Nameki,
\textit{``Perturbative study of the transfer matrix on the string worldsheet in
  $AdS_5 \times S^5$''},
\href{http://arXiv.org/abs/arXiv:0706.1525 [hep-th]}{\texttt{arXiv:0706.1525
  [hep-th]}}.
%
%%CITATION = HEP-TH 9805028;%%
\bibitem{Metsaev:1998it}
R.~R.~Metsaev and A.~A.~Tseytlin,
\textit{``Type IIB superstring action in $AdS_5 \times S^5$ background''},
\textsf{Nucl.~Phys.~B533,~109~(1998)},
\href{http://arXiv.org/abs/hep-th/9805028}{\texttt{hep-th/9805028}}.
%
%%CITATION = HEP-TH 0612079;%%
\bibitem{Maldacena:2006rv}
J.~Maldacena and I.~Swanson,
\textit{``Connecting giant magnons to the pp-wave: An interpolating limit of
  $AdS_5 \times S^5$''},
\href{http://arXiv.org/abs/hep-th/0612079}{\texttt{hep-th/0612079}}.
%
%%CITATION = ARXIV:0707.1676;%%
\bibitem{Benvenuti:2007qt}
S.~Benvenuti and E.~Tonni,
\textit{``Near-flat space limit and Einstein manifolds''},
\href{http://arXiv.org/abs/arXiv:0707.1676 [hep-th]}{\texttt{arXiv:0707.1676
  [hep-th]}}.
%
%%CITATION = NUPHA,B174,166;%%
\bibitem{Parke:1980ki}
S.~J.~Parke,
\textit{``Absence of particle production and factorization of the S-matrix in
  $(1+1)$-dimensional models''},
\textsf{Nucl.~Phys.~B174,~166~(1980)}.
%
%%CITATION = HEP-TH/0701281;%%
\bibitem{Torrielli:2007mc}
A.~Torrielli,
\textit{``Classical r-matrix of the $su(2|2)$ SYM spin-chain''},
\textsf{Phys.~Rev.~D75,~105020~(2007)},
\href{http://arXiv.org/abs/hep-th/0701281}{\texttt{hep-th/0701281}}.
%
%%CITATION = ARXIV:0705.1071;%%
\bibitem{Heckenberger}
I.~Heckenberger, F.~Spill, A.~Torrielli and H.~Yamane,
\textit{``Drinfeld second realization of the quantum affine superalgebras of
  $D^{(1)}(2,1;x)$ via the Weyl groupoid''},
\href{http://arXiv.org/abs/arXiv:0705.1071 [math.QA]}{\texttt{arXiv:0705.1071
  [math.QA]}}.
%
%%CITATION = ARXIV:0706.0884;%%
\bibitem{Moriyama:2007jt}
S.~Moriyama and A.~Torrielli,
\textit{``A Yangian Double for the AdS/CFT Classical r-matrix''},
\href{http://arXiv.org/abs/arXiv:0706.0884 [hep-th]}{\texttt{arXiv:0706.0884
  [hep-th]}}.
%
%%CITATION = NLIN/0610017;%%
\bibitem{Beisert:2006qh}
N.~Beisert,
\textit{``The Analytic Bethe Ansatz for a Chain with Centrally Extended
  $su(2|2)$ Symmetry''},
\textsf{J.~Stat.~Mech.~0701,~P017~(2007)},
\href{http://arXiv.org/abs/nlin/0610017}{\texttt{nlin/0610017}}.
%
%%CITATION = HEP-TH/0603008;%%
\bibitem{Frolov:2006cc}
S.~Frolov, J.~Plefka and M.~Zamaklar,
\textit{``The $AdS_5 \times S^5$ superstring in light-cone gauge and its Bethe
  equations''},
\textsf{J.~Phys.~A39,~13037~(2006)},
\href{http://arXiv.org/abs/hep-th/0603008}{\texttt{hep-th/0603008}}.
%
%%CITATION = HEP-TH/0603039;%%
\bibitem{Klose:2006dd}
T.~Klose and K.~Zarembo,
\textit{``Bethe ansatz in stringy sigma models''},
\textsf{J.~Stat.~Mech.~0605,~P006~(2006)},
\href{http://arXiv.org/abs/hep-th/0603039}{\texttt{hep-th/0603039}}.
%
%%CITATION = HEP-TH/0609157;%%
\bibitem{Arutyunov:2006ak}
G.~Arutyunov, S.~Frolov, J.~Plefka and M.~Zamaklar,
\textit{``The off-shell symmetry algebra of the light-cone $AdS_5 \times S^5$
  superstring''},
\textsf{J.~Phys.~A40,~3583~(2007)},
\href{http://arXiv.org/abs/hep-th/0609157}{\texttt{hep-th/0609157}}.
%
\bibitem{Kallen:1965}
G.~K{\"a}llen and J.~Toll,
\textit{``Special Class of Feynman Integrals in Two-Dimensional Space-Time''},
\textsf{J.Math.Phys~6,~299~(1965)}.
%
%%CITATION = PHRVA,D11,838;%%
\bibitem{Thacker:1974kv}
H.~B.~Thacker,
\textit{``Bethe's Hypothesis and Feynman Diagrams: Exact Calculation of a Three
  Body Scattering Amplitude by Perturbation Theory''},
\textsf{Phys.~Rev.~D11,~838~(1975)}.
%
%%CITATION = PHRVA,D14,3508;%%
\bibitem{Thacker:1976vp}
H.~B.~Thacker,
\textit{``Many Body Scattering Processes in a One-Dimensional Boson System''},
\textsf{Phys.~Rev.~D14,~3508~(1976)}.
%
%%CITATION = HEP-TH/9412039;%%
\bibitem{Grabowski:1994rb}
M.~P.~Grabowski and P.~Mathieu,
\textit{``Integrability test for spin chains''},
\textsf{J.~Phys.~A28,~4777~(1995)},
\href{http://arXiv.org/abs/hep-th/9412039}{\texttt{hep-th/9412039}}.
%
%%CITATION = SVMDA,32,254;%%
\bibitem{Drinfeld:1985rx}
V.~G.~Drinfeld,
\textit{``Hopf algebras and the quantum Yang-Baxter equation''},
\textsf{Sov.~Math.~Dokl.~32,~254~(1985)}.
%
\end{thebibliography}

\end{document}